\documentclass[pra,aps,superscriptaddress,twocolumn]{revtex4}

\usepackage[english]{babel}
\usepackage{mathtools,amsmath,amsfonts,amssymb}
\usepackage{graphicx}
\usepackage[hidelinks]{hyperref}
\usepackage{float}
\usepackage{bm,dsfont}
\usepackage[dvipsnames]{xcolor}
\usepackage{color}
\usepackage[normalem]{ulem}
\usepackage{bbm}
\usepackage{placeins}
\usepackage{csquotes}
\usepackage{braket}
\usepackage{stackengine}\stackMath

\newcommand{\dicke}[2]{\ket{D^{#1}_{#2}}}
\newcommand{\W}[1]{\ket{W_{#1}}}

\begin{document}
\title{$W$- and Dicke-state engineering using optimal global control in nearest-neighbor coupled ring-shaped qubit arrays}

\author{Andrea Muratori}
\affiliation{Dipartimento di Fisica e Astronomia, Università di Bologna, 40127 Bologna, Italy}

\author{Vladimir M. Stojanovi\'c}
\affiliation{Institut für Angewandte Physik, Technical University of Darmstadt, D-64289 Darmstadt, Germany}

\author{Eloisa Cuestas}
\affiliation{Forschungszentrum Jülich GmbH, Peter Grünberg Institute, Quantum Control (PGI-8), 52425 Jülich, Germany}
\affiliation{Quantum Systems Unit, Okinawa Institute of Science and Technology Graduate University, Onna, Okinawa 904-0495, Japan}

\author{Tommaso Calarco}
\affiliation{Dipartimento di Fisica e Astronomia, Università di Bologna, 40127 Bologna, Italy}
\affiliation{Forschungszentrum Jülich GmbH, Peter Grünberg Institute, Quantum Control (PGI-8), 52425 Jülich, Germany}
\affiliation{Institute for Theoretical Physics, University of Cologne, Zülpicher Straße 77, 50937 Cologne, Germany}

\author{Felix Motzoi}
\affiliation{Forschungszentrum Jülich GmbH, Peter Grünberg Institute, Quantum Control (PGI-8), 52425 Jülich, Germany}
\affiliation{Institute for Theoretical Physics, University of Cologne, Zülpicher Straße 77, 50937 Cologne, Germany}

\date{\today}

\begin{abstract}
    Motivated by a compelling need for time-efficient and robust schemes for quantum-state engineering in systems of neutral atoms in optical tweezers, we consider a ring-shaped array of qubits with nearest-neighbor Ising-type ($zz$) coupling and transverse ($x$ and $y$) global control fields. This system to a large extent mimics -- outside of the Rydberg-blockade regime -- a circular array of neutral atoms interacting through van-der-Waals type interaction. We investigate the preparation of $W$ and Dicke states in this system starting from the default initial state $|00\ldots 0\rangle$ using two different optimal-control approaches: (i) NMR-like pulse sequence, which consists of instantaneous (delta-shaped) control- and Ising-interaction pulses, and (ii) time-dependent control scheme, which entails shaped control pulses in the presence of always-on Ising interaction between adjacent qubits. By making use of the underlying dihedral symmetry of this system -- which allows one to use a symmetry-adapted computational basis with $\mathcal{O}(2^N / N)$ states in an $N$-qubit system -- and utilizing advanced global-optimization methods, we find optimal sequences of pulses for realizing $W$ and Dicke states within both approaches. In addition, we demonstrate robustness of these sequences against unavoidable control errors. Finally, using typical values of parameters in realistic Rydberg-atom systems, we show that our control schemes enable the preparation of the desired multiqubit states on time scales much shorter than the relevant coherence times of those systems.
\end{abstract}

\maketitle

\section{Introduction}
Owing to significant advances of the optical-tweezer technology and the related methods of manipulation and control, recent years have witnessed tantalizing progress in scaling up neutral-atom qubit arrays~\cite{Endres+:16,Barredo+:18,Schymik+:20}. These experimental developments, along with the ensuing achievements in neutral-atom quantum computing (QC) and analog simulation~\cite{Browaeys+Lahaye:20}, has spawned ever-increasing interest in quantum-state engineering in neutral-atom arrays. This has led to the experimental realization of interesting quantum states of neutral-atom ensembles, such as Schr\"{o}dinger cat states~\cite{Omran+:19}. In addition, a number of theoretical proposals for engineering maximally-entangled multiqubit states of $W$ and Greenberger-Horne-Zeilinger (GHZ) types~\cite{Zheng+:20,Stojanovic:21,Haase+:21, Haase+:22,Nauth+Stojanovic:22, Shao+:23,Jin+Jing:25}, Knill-Laflamme-Milburn states~\cite{Zheng+:21}, cluster-~\cite{Crescimanna+:23} and graph states~\cite{Li+:24}, spin-squeezed states~\cite{Carrera+:25}, as well as quantum superpositions of current states~\cite{Perciavalle+:23,Perciavalle+:24}, have been put forward in recent years. While some of these proposals assume the presence of the resonant dipole-dipole (dipolar) interaction between Rydberg atoms~\cite{Stojanovic:21,Perciavalle+:23, Perciavalle+:24, Li+:24}, the remaining ones correspond to systems with the off-resonant dipole-dipole (van-der-Waals) interaction~\cite{Zheng+:20,Haase+:21,Haase+:22,Nauth+Stojanovic:22,Shao+:23,Zheng+:21, Crescimanna+:23,Jin+Jing:25,Carrera+:25}.

Prompted by a pressing need to design time-efficient and robust preparation of highly-entangled multiqubit states in Rydberg-atom systems, in this paper we carry out a systematic, optimal-control-based~\cite{Koch+:22,Mueller+:22,Ansel+:24} investigation of the preparation of $W$ and Dicke states in such systems. To be more specific, we investigate the preparation of the latter states in circular (ring-shaped) qubit arrays in which adjacent qubits are coupled through Ising-type ($zz$) interaction, while at the same time being subject to global transverse ($x$ and $y$) control fields. This last physical model to a large extent mimics the physics of circular arrays of ground-Rydberg ($gr$) type neutral-atom-based qubits~\cite{Haase+:21,Nauth+Stojanovic:22} interacting through van-der-Waals-type interaction outside of the Rydberg-blockade regime; the fact that Ising-type coupling between qubits in our model is limited to nearest neighbors only reflects the fast spatial decay ($\propto r^{-6}$) of the van-der-Waals interaction. While the Rydberg-blockade regime provides the natural setting for spontaneous creation of $W$-type entanglement -- namely, in the form of a collective state of $N$ neutral atoms with a single Rydberg excitation shared among all of them -- our study aims to provide an easy-to-implement recipe for realizing the same type of multipartite entanglement outside of this last regime. 

It is worthwhile noting that pulse-level (analog) schemes for engineering permutationally-invariant $W$, GHZ, and Dicke states in the case of all-to-all Ising-coupled qubits with transverse global control have already been proposed in recent years~\cite{Chen+:17,Stojanovic+Nauth:22,Stojanovic+Nauth:23,
Koutromanos+:24,Stefanescu+:25,deLima+:25}. When dealing with the latter type of models, which are also invariant under the full permutation group $S_N$, one can restrict oneself to the permutationally-invariant subspace (symmetric sector) of the $N$-qubit Hilbert space, whose dimension ($N+1$) is linear in the number of qubits~\cite{Stojanovic+Nauth:23}. By contrast to these recent investigations, in the present work we address the more challenging case of a ring-shaped qubit array with nearest-neighbor (rather than all-to-all) Ising-type interaction between qubits, a system characterized by the dihedral symmetry (i.e. the symmetry group $D_N$ of a regular polygon with $N$ sides)~\cite{Lyons+:22}. In this case the symmetry-adapted basis of the $N$-qubit Hilbert space has the dimension $\mathcal{O}(2^N / N )$, which is only by a factor of $N$ smaller than that of a generic $N$-qubit system. Therefore, the problem of engineering complex, highly-entangled multiqubit states in such a system presents a highly nontrivial computational challenge, which we aim to address in what follows by employing the advanced toolbox of quantum optimal control~\cite{Koch+:22,Mueller+:22,Ansel+:24}.

We explore the preparation of $W$ and Dicke states in ring-shaped qubit arrays within two different control scenarios. We first investigate an NMR-type scheme in which global-control pulses are assumed to be instantaneous (delta-shaped)~\cite{Vandersypen+Chuang:05}; such pulses are equivalent to global qubit rotations and were utilized in the past for 
generating~\cite{Das+:15,Stojanovic+Nauth:23} and preserving~\cite{Tanamoto+:12,Tanamoto+:13} 
highly-entangled multiqubit states, as well as for deterministic conversions between maximally-entangled ones~\cite{Stojanovic+Nauth:22,Zhou+:25}. The ensuing state-preparation scheme then has the form of a pulse sequence that consists of global qubit rotations and Ising-interaction pulses. We then complement this last NMR-type approach by exploring a scenario that entails time-dependent control pulses, i.e. continuous-wave global control; these pulses are parametrized using a suitably chosen basis of time-dependent functions~\cite{motzoi2011optimal}.

We show that our two control schemes allow time-efficient and robust preparation of $W$ and Dicke states. We demonstrate that -- for typical values of parameters characterizing Rydberg-atom systems -- the obtained state-preparation times are considerably shorter than typical radiative lifetimes of Rydberg states. This is a strong indication for the potential practical usefulness of our proposed state-preparation schemes for realistic neutral-atom qubit arrays.

The remainder of this paper is organized as follows. We introduce the model of globally-controlled 
Ising-coupled qubits in a circular topology in Sec.~\ref{ModelAndTargets}, where we also substantiate the relevance of this model in the context of neutral-atom systems. We also provide a brief review of the two classes of entangled multiqubit states to be discussed in this work. Finally,
we also provide a short discussion of the symmetry- and 
controllability aspects of the problem under consideration.
In Sec.~\ref{sec:nmr} we describe the NMR-type approach to the state-engineering problem at hand, along with a discussion of the results obtained using this approach. The alternative, time-dependent approach -- along with the ensuing results for the state-preparation problem under consideration -- is presented in Sec.~\ref{sec:tdep}. Section~\ref{sec:implementRydberg} 
is devoted to the discussion of possible implementation of our proposed $W$- and Dicke state-engineering in realistic 
neutral-atom arrays in optical tweezers.
We conclude, with some general remarks and outlook, in Sec.~\ref{sec:SumConcl}. Some 
symmetry-related considerations are relegated
to Appendix~\ref{app:more-symmetries}, while certain technical details pertaining to the numerical implementations of the two optimal-control approaches utilized in this work are provided in Appendices~\ref{app:CodeTechn}
and \ref{app:CodeTimeDep}.

\section{Model and target multiqubit states} \label{ModelAndTargets} 
To set the stage for further considerations, in what follows 
we introduce the model under consideration and the target multiqubit states. To begin with,
in Sec.~\ref{ModelHam} we specify the full Hamiltonian of the system and explain its relevance in the context of neutral-atom qubit arrays. The two relevant classes of entangled multiqubit states (Dicke and $W$)
are briefly introduced in Sec.~\ref{TargetStates}. 
In Sec.~\ref{SymmSystem} we discuss the dihedral symmetry of the system at hand
and its implications. Finally, Sec.~\ref{Controllability} is devoted to a short discussion of the controllability aspects of the problem under consideration.

\subsection{Model Hamiltonian} \label{ModelHam}
The model under consideration
is that of a ring-shaped qubit array (for illustration, see Fig.~\ref{fig:QubitRing}
below) 
that features nearest-neighbor Ising-type ($zz$) interaction and is at the same time subject to global transverse ($x$ and $y$) Zeeman-like control fields. In such a circular array, the qubits are positioned in the vertices of a regular polygon.

The assumed physical realization of qubits 
in this system corresponds to a $gr$-type
Rydberg-atom qubit, where the role of the 
logical $|0\rangle$ and $|1\rangle$ qubit 
states are played by the atomic ground state
$|g\rangle$ and a high-lying Rydberg state 
$|r\rangle$, respectively. When expressed in frequency units, the typical energy splitting in $gr$-type qubits is in the range between $900$ and 
$1500$\:THz~\cite{Haase+:21}, 
where the actual value depends on
the choice of the atomic species and Rydberg states used. Therefore,
practical manipulations of $gr$-type qubits require either an ultraviolet laser or a combination of visible and infrared lasers in a ladder configuration. In addition to their fairly straightforward initialization, manipulation, and measurement -- compared to other types of neutral-atom qubits -- such qubits allow fast, high-fidelity entangling operations~\cite{Haase+:21}.

\begin{figure}[t!]
\includegraphics[clip,width=0.95\linewidth]{IsingStatePrepDickeW_Fig1.pdf}
\caption{\label{fig:QubitRing}Pictorial
illustration of a ring-shaped qubit array,
with nearest-neighbor Ising-type coupling of strength $J$ 
and transverse global control fields $B_x(t)$ and $B_y(t)$. 
The qubits in this system are assumed to be $gr$-type 
neutral-atom qubits.}
\end{figure}

The total Hamiltonian governing this model system is given by
\begin{equation}
    H(t)= H_{ZZ}+H_C (t) \:,
    \label{eq:ham}
\end{equation}
where $H_{ZZ}$ is the (time-independent) Ising-type drift Hamiltonian, 
\begin{equation} \label{DriftHam}
H_{ZZ} = J\sum_{n=1}^N\sigma_n^z
\sigma_{n+1}^z \:,
\end{equation}
where $J$ is the Ising-coupling strength, while $H_C (t)$ is the (time-dependent) global-control part that describes the $x$- and 
$y$ control fields:
\begin{equation}\label{ControlHam}
\begin{split}
    H_C (t) &= B_x(t)\sum_{n=1}^N \sigma_n^x + 
    B_y(t)\sum_{n=1}^N \sigma_n^y\\
    &= B_x(t)H_x+B_y(t)H_y\:.
\end{split}
\end{equation}
As a consequence of the assumed ring-shaped topology of the qubit array, the sums over $n$ in  Eqs.~\eqref{DriftHam} and \eqref{ControlHam}
run over the $N$ qubits in the array and periodic boundary conditions are imposed, i.e. $\sigma_{N+1}^{\mu}\equiv\sigma_{1}^{\mu}\:\: (\mu =x,y,z)$.

In what follows, we set $\hbar=1$.
All the energies in the problem under consideration will hereafter be
expressed in units of the Ising-coupling strength $J$. Accordingly, all the relevant timescales will be expressed in units of $J^{-1}$.

\subsection{State preparation as a 
quantum-control problem and 
target multiqubit states} \label{TargetStates}
Our primary objective in what follows is to
demonstrate the feasibility of engineering certain highly-entangled multiqubit states in the system of Ising-coupled qubits described in
Sec.~\ref{ModelHam}. 

In all cases considered, we will assume the default state 
$|00\ldots 0\rangle$ to be the initial state $|\psi(t=0)\rangle$ of our $N$-qubit system. We will aim to find the appropriate time dependence of the global control fields $B_x(t)$ and $B_y(t)$ that allows us to prepare a desired $N$-qubit state $\ket{\psi_{\textrm{target}}}$ at time $t=t_f$. Moreover, in the spirit of quantum optimal control, we will aim to find the shortest possible time $t_f$ needed to prepare the desired $N$-qubit state. The figure of merit quantifying the envisioned control-based state-preparation process is the target-state fidelity 
\begin{equation}\label{eq:fidelity}
{\cal F}_{t=t_f}=|\langle\psi_{\textrm{target}}
\ket{\psi(t=t_f)}|^2 \:,
\end{equation}
i.e. the module squared of the overlap between the actual state of the system at time $t=t_f$
\begin{equation}
\ket{\psi(t=t_f)}=U(t=t_f)\ket{\psi(t=0)} \:,
\end{equation}
where $U(t)$ is the time-evolution operator of the system, and the target state 
$\ket{\psi_{\textrm{target}}}$.

In what follows, we briefly introduce the two families of highly-entangled, permutationally-symmetric, multiqubit states -- Dicke and $W$ states -- whose preparation starting from the state $|00\ldots 0\rangle$ we are concerned with in this work.

Dicke state $\dicke Nk$ is given by the equal-weight superposition of all $N$-qubit product states with exactly $k$ excitations, i.e.~the states corresponding to the bitstrings of length $N$ with $k$ occurrences of $1$ and $N-k$ occurrences of $0$ (i.e.~of Hamming weight $k$). Dicke states can be formally written in the form
\begin{equation}\label{eq:dicke}
\dicke Nk = \binom{N}{k}^{-1/2}\sum_{P}
P\{\ket{1}^{\otimes k}\ket{0}^{\otimes{N-k}}\}
\: .
\end{equation}
where in the sum on the right-hand-side of 
the last equation $P$ stands for an arbitrary  permutation of the set $\{1,2,\ldots,N\}$. 
For example, the states $\dicke32$
and $\dicke42$ respectively read as follows:
\begin{eqnarray} \label{Dicke32and42}
\dicke32 &=& \frac{1}{\sqrt{3}}\:(|110\rangle
+|101\rangle+|011\rangle) \:,\\
\dicke42 &=& \frac{1}{\sqrt{6}}\:(|1100\rangle
+|1010\rangle+|1001\rangle \nonumber\\
& &+ |0110\rangle + |0101\rangle + |0011\rangle) \:.\nonumber
\end{eqnarray}

As a notable special case of the family of Dicke states, we will also focus our attention in the following on the preparation of $W$ states. The $W$ state of an $N$-qubit system is the maximally-entangled state given by an equal-weight superposition of all $N$-qubit product states with a single excitation (i.e. Hamming-weight-$1$ states), i.e.
\begin{equation} \label{eq:w}
\W N=\frac{1}{\sqrt N}(|10\ldots 0\rangle + 
|01\ldots 0\rangle +\ldots + |00\ldots 1\rangle)\: .
\end{equation}
In the $N=3$ case, for instance, it corresponds to one ground state of the frustrated antiferromagnetic Ising model:
\begin{equation}
    \W3=\frac1{\sqrt{3}}(\ket{001}+\ket{010}+\ket{100})\ .
\end{equation}

Because of their favorable properties -- especially their robustness to particle loss~\cite{Neven+:18}
-- both Dicke and $W$ states hold promise for quantum-technology applications~\cite{Stojanovic+Nauth:23}. This has prompted attempts to realize 
$W$-\cite{Stojanovic:21,Kang+:16,Fang+:19,
StojanovicPRL:20,Zhang+:23} and Dicke 
states~\cite{Stockton+:04,Xiao+:07,Hume+:09,
Wieczorek+:09,Lamata+:13} in various physical platforms. 

\subsection{Dihedral symmetry of the system} \label{SymmSystem}
Looking both at the Hamiltonian of our system in Eqs.~(\ref{eq:ham},\ref{DriftHam},\ref{ControlHam}) and at the states that we are targeting in our preparation protocol in Eqs.~(\ref{eq:dicke},\ref{eq:w}), we can notice that they share a symmetry on which we will heavily rely for our computations.
The relevant symmetry group is the dihedral group $D_N$, the symmetry group of a polygon with $N$ vertices. This group has $2N$ elements -- $N$ rotations (i.e. cyclic permutations of the polygon's vertices) and $N$ reflections around symmetry axes.
The elements of $D_N$ thus act on a generic state $\ket{a_1\dots a_N}$ ($a_i=0,1$) as
\begin{equation}
\begin{split}
    C_N^s\ket{a_1\dots a_N}&=\ket{a_{s+1}\dots a_N a_1 \dots a_s}\\
    C_N^{(s)}\ket{a_1\dots a_N}&=R\ C_N^{s}\ket{a_1\dots a_N}\\&=\ket{a_{s}\dots a_1 a_N \dots a_{s+1}}
\end{split}
\end{equation}
where $C_N^s$ is a cyclic permutation of $s$ sites and $R$ is the reversing of a string.

The basis of the $D_N$-invariant subspace of the total Hilbert space  can readily be found by symmetrizing the computational basis under the group action. In particular, the $D_N$-invariant subspace will be spanned by orthonormal vectors given by the equal superposition of computational states belonging to the orbits of $D_N$ acting on the computational basis itself. Denoting by ${\cal H}_{D_N}$ such a subspace of the full Hilbert space ${\cal H}$ and by ${\rm CB}$ the computational basis, we have
\begin{equation}
\begin{split}
        {\cal H}_{D_N} &= {\rm span} \left\{\sum_{y\in{\rm orbit}(x)}\frac{y}{|{\rm orbit}(x)|},\ x\in{\rm CB}\right\}\\
        &= {\rm span}\left\{{\rm CB}_{D_N}\right\} \:,
\end{split}
\end{equation}
with
\begin{equation}
    {\rm orbit}(x) = \{g(x)\}_{g\in D_N} \ .
\end{equation}
The definition of the basis-changing (and projecting) matrix $P$ from $\cal H$ to ${\cal H}_{D_N}$ follows straightforwardly as the one having the spanning vectors of ${\cal H}_{D_N}$ as columns. Therefore, the representation of the new matrices in the $D_N$-symmetrical subspace is given by $\tilde H_{x,y,ZZ}=P^\dag H_{x,y,ZZ} P$. 
The dimension of the subspace ${\cal H}_{D_N}$ is then $\mathcal{O}(2^N/N)$; even if this is not an exponential dimensionality reduction from the original $2^N$-dimensional Hilbert space of the $N$-qubit system (as would be, for instance, the case for the 
full permutational symmetry~\cite{Stojanovic+Nauth:23}), it still provides a few orders of magnitude of computational speedup~\cite{wang_accelerating_2023}.

Further symmetries of the system could in principle be taken into account (for details, see Appendix~\ref{app:more-symmetries}). For the sake of computations, however, we will not resort to them, sticking only to the generic dihedral one.

\subsection{State-to-state controllability of the system} \label{Controllability}
As a prelude to designing optimal-control-based 
schemes for engineering $W$- and Dicke states in 
the system at hand (see Secs.~\ref{sec:nmr} and \ref{sec:tdep} below), it is pertinent to briefly
discuss the controllability-related aspects 
of the quantum state-engineering problem under consideration. The controllability criteria for finite-dimensional quantum systems, framed using Lie-algebraic concepts, are existence theorems~\cite{D'AlessandroBook} that -- if their conditions of validity are fulfilled -- guarantee that time-dependent control fields can be found so as to be able to carry out a given state-~\cite{Stojanovic+Nauth:23} or operator(gate)-related~\cite{Stojanovic:19}
control task. They do not, however, specify the appropriate time dependence of those fields; finding the actual time-dependent fields usually entails optimal-control methods, as is the case in the present work.

The controllability-related aspects of interacting spin-$1/2$ chains (qubit arrays) were discussed in  considerable detail in the past~\cite{D'AlessandroBook}. It has long since been known that for the complete (operator) controllability -- equivalent to universal QC -- of a spin-$1/2$ chain with nearest-neighbor Ising interaction it is necessary to have two mutually noncommuting controls acting on each spin in the chain~\cite{D'AlessandroBook};
the same is true of the $XY$ interaction,
while only Heisenberg-type interactions require a smaller degree of control for complete controllability~\cite{Heule+:10,Heule+:11}. 
Based on this established general result, it is clear that an $N$-qubit 
array with Ising-type coupling between adjacent qubits and global control fields in the $x$ and $y$ directions [cf.~Eqs.~(\ref{DriftHam},\ref{ControlHam}) above],
a system of the kind considered in the present work,
is not completely controllable. In other words, its dynamical Lie algebra $\mathcal{L}_d$, generated by $H_{ZZ}$, $H_{x}$, and $H_{y}$, is not isomorphic with the full $u(2^N)$ or $su(2^N)$ algebra, but with their proper Lie subalgebra.

The lack of complete controllability of the system 
at hand immediately implies that this system is also
not state-to-state controllable for a generic choice of the initial- and final states in its dynamical evolution. However, the system is state-to-state controllable in the physical scenario of interest 
for our anticipated control schemes, 
in which both the initial 
state $|00\ldots 0\rangle$ of the qubit array and its final ($W$- or Dicke) state are
permutationally symmetric. 
This statement can be reduced to another established result pertaining to controllability of symmetric spin networks;
namely, it is known that a system of 
spin-$1/2$ particles described by a permutationally
invariant Hamiltonian is state-to-state controllable for an arbitrary pair of permutationally invariant
initial and final states~\cite{Albertini+DAlessandro:18}. 
If such a system is subject to global transverse ($x$ and $y$)
control fields, like those in Eq.~\eqref{ControlHam} above,
this last result implies that a time dependence of the global control fields 
can be found such that an arbitrary permutationally invariant final state can be reached through 
an evolution of such a system 
that starts from an initial state with that same symmetry. It is worthwhile recalling 
that the system at hand does not have the full permutational symmetry $S_N$ but only the dihedral symmetry $D_N$ [cf. Sec.~\ref{SymmSystem}]. Yet, given that $D_N$ is a proper subgroup of $S_N$ and that the $S_N$-invariant initial and final states [$|00\ldots 0\rangle$ and $|W_N\rangle$ (or $|D^{N}_k\rangle$), respectively] are also $D_N$-invariant, the previously mentioned result on the state-to-state controllability that holds for $S_N$-symmetric states can be generalized {\em mutatis mutandis} to the $D_N$-symmetric ones in the problem at hand. 

The established controllability result implies -- for an $N$-qubit array -- that 
time-dependent global control fields $B_x(t)$ and $B_y(t)$ in Eq.~\eqref{ControlHam} can be found so as to realize the state $|W_N\rangle$, or any Dicke state $|D_k^N\rangle$,
starting from the state $|00\ldots 0\rangle$ at $t=0$. 
To obtain the time-dependence of these fields that allows one to 
engineer $W$- and Dicke states in the shortest possible time,
we employ in what follows two different optimal-control schemes.

\section{$W$- and Dicke-state preparation with NMR-like pulses}\label{sec:nmr}
In this section, we discuss the generation of Dicke and $W$ states using an NMR-like control approach based on sequences with a finite number of pulses. The general aspects 
of this approach are first discussed in Sec.~\ref{sec:nmr-problem}. This is followed 
by a discussion, in Sec.~\ref{sec:nmr-results}, of the results that this approach yields in the 
state-engineering problem under 
consideration.

\subsection{NMR-type pulse sequence for 
state preparation and its optimization}\label{sec:nmr-problem}
The physical scenario involving NMR-like pulse sequences corresponds to experiments in which very strong electromagnetic fields are intermittently coupled to the atoms for a very short amount of time.
Because of this, the system dynamics can be well approximated by instantaneous (delta-shaped) global-control pulses, alternated with (constant) Ising-interaction pulses of finite duration. Therefore,
the tunable parameters in an ensemble of neutral-atom
based qubits under such physical circumstances are the global-field strengths corresponding to each of the delta-shaped pulses, as well as the durations of the interaction pulses.

The instantaneous global control pulses are equivalent to global qubit rotations. For a delta-shaped pulse
at $t=T_p$, the global transverse control field $\mathbf{B}(t)\equiv [B_x(t), B_y(t), 0]^{\textrm{T}}$ has the form 
$\mathbf{B}(t)=\bm{\alpha}
\delta(t-T_p)$,
where $\bm{\alpha}$ is a vector in the $x-y$ plane; denoting its norm by $\alpha$ and its corresponding polar angle by $\phi$, this vector is given by $\bm{\alpha}=(\alpha \cos\phi, \alpha \sin\phi, 0)^{\textrm{T}}$. 
Because in this case the $x$- 
and $y$- control fields have the same time-dependence, the control  Hamiltonian $H_C(t)$ belongs to the class of time-dependent Hamiltonians that commute with themselves at different times, i.e.~satisfy the condition $[H_C(t),H_C(t')]=0$. Consequently, the corresponding 
time-evolution operator adopts the form $\exp[-i\int_{t_i}^{t_f}H_C (t)dt]$, with $t_i$ and $t_f$ being the initial and final evolution times, respectively. This time-evolution operator evaluates to the exponential of a linear combination of the operators 
$H_{x}$ and $H_{y}$ [cf. Eq.~\eqref{ControlHam}], i.e.
\begin{equation}\label{def:Uxy}
U_{xy}(\alpha,\phi)=
\exp\left[-i\alpha (H_x\cos\phi+H_y\sin\phi)\right] \:.
\end{equation}

Given the form of $H_{x}$ and $H_{y}$ [cf. Eq.~\eqref{ControlHam}],
$U_{xy}(\alpha,\phi)$ is also an exponential of a linear combination 
of the Pauli operators $\sigma^{x}_n$
and $\sigma^{y}_n$ for all the qubits
($n=1,2,\ldots,N$). Recalling that single-qubit rotations are described by the operators $R_{\mathbf{\hat{n}}}(2\theta)\equiv\exp[-i\theta (\mathbf{\hat{n}}\cdot\bm{\sigma})]$ [where $\theta$ is the half-angle of rotation, 
$\mathbf{\hat{n}}$ a unit vector defining 
the rotation axis, and $\bm{\sigma}\equiv(\sigma^x,\sigma^y,\sigma^z)^{\textrm{T}}$ the vector of 
single-qubit Pauli operators], this last statement makes manifest the equivalence between global delta-shaped control pulses and global qubit rotations. Therefore,
a delta-shaped control pulse described by the vector $\boldsymbol{\alpha}$ 
amounts to a global qubit rotation through an angle of $2\alpha$ around the axis whose direction coincides with that of the unit vector $\mathbf{\hat{n}}
\equiv(\cos\phi,\sin\phi,0)^{\textrm{T}}$ in the $x-y$ plane. For an actual evaluation of 
$U_{xy}(\alpha,\phi)$ [cf. Eq.~\eqref{def:Uxy}], it is pertinent to make use of 
the well-known Euler identity  
\begin{equation}\label{PauliExpIdentity}
\exp[-i\theta (\mathbf{\hat{n}}\cdot\bm{\sigma})]= 
\cos\theta \mathbbm{1}_{2}- i\sin\theta\:(\mathbf{\hat{n}}\cdot \bm{\sigma}) 
\end{equation}
for single-qubit rotation operators.

Let us consider an NMR-type pulse sequence with the aim of engineering our target $W$ and Dicke states. Let us assume that this pulse sequence consists 
of a certain number of delta-shaped global-control pulses
with the parameters 
$\alpha_j$ and $\phi_j$, as well as Ising-interaction pulses 
with the dimensionless 
(expressed in units of $J^{-1}$) durations $\xi_j$ ($j=1,2,\ldots$). It can be established that such a pulse sequence should consist of $2M+1$ time steps, of which $M+1$ are control pulses and $M$ their Ising-interaction counterparts. The time-evolution operator representing the entire pulse sequence can be written as
\begin{equation}\label{defTotalU}
U=U_{xy}(\alpha_{M+1},\phi_{M+1}) 
\prod_{j=M}^1 U_{ZZ}(\xi_j)\: 
U_{xy} (\alpha_j,\phi_j) \:,
\end{equation}
where $U_{ZZ}(\xi_j)=\exp(-i\xi_j H_{ZZ})$
is the time-evolution operator corresponding to $j$-th interaction pulse ($j=1,2,\ldots, M$). In connection with the form of Eq.~\eqref{defTotalU}, it is pertinent to note that -- given that our initial state is $\ket0^{\otimes N}$ and that all of our target ($W$ and Dicke) states are superpositions of eigenstates of the $z$-parity operator
$Z_P\equiv\sigma^z_1\ldots \sigma^z_N$ -- the 
first- and final pulses 
must be global-control 
(qubit rotation) pulses [cf. Eq.~\eqref{def:Uxy}].

In order to maximize the target-state fidelity [cf. Eq.~\eqref{eq:fidelity}], we have to carry out its optimization
over the parameters $\{\alpha_j,\phi_j,\xi_j| 
j=1,2,\ldots,M\}$, as well 
as over the parameters 
$\{\alpha_{M+1},\phi_{M+1}\}$ of the final global-control
pulse [cf. Eq.~\eqref{defTotalU}]. Moreover, the value of $M$ 
can be optimized as well, 
for example through the bisection method, with convergence reached at the minimal $M$ such that a given threshold value of the figure of merit $\mathcal{F}_{t=t_f}$ is exceeded. Because of the large number of optimization parameters and the complexity of the landscape, the global optimization of the target-state fidelity is performed by making use of the multistart-based global random initialization method~\cite{TornZilinskasBook}, which is 
particularly suitable in cases where the 
objective function has a multitude of nearly degenerate local minima~\cite{Stojanovic+Vanevic:08}. In the
problem at hand, tens of thousands of initial guesses are evaluated, followed by local searches for the minima of the target-state infidelity $1-\mathcal{F}_{t=t_f}$ around a smaller number of the best initial guesses, i.e.~those that correspond to the smallest values of the objective function that is to be minimized (in this case, $1-\mathcal{F}_{t=t_f}$). In this manner, we collected results from thousands of local optimizations and performed statistical analysis on the best ones. More technical details 
about code implementation can be found in Appendix~\ref{app:CodeTechn}.

\subsection{Pulse sequences for $W$- and Dicke-state preparation: results and discussion}\label{sec:nmr-results}
We present here the results obtained for the preparation of $W$ and Dicke states in the NMR scheme with up to $N=9$ qubits.
Throughout this Section, we will indicate with just $\cal F$ the target-state fidelity ${\cal F}_{t=t_f}.$

\begin{figure}
    \centering
    \includegraphics[width=0.95\columnwidth]{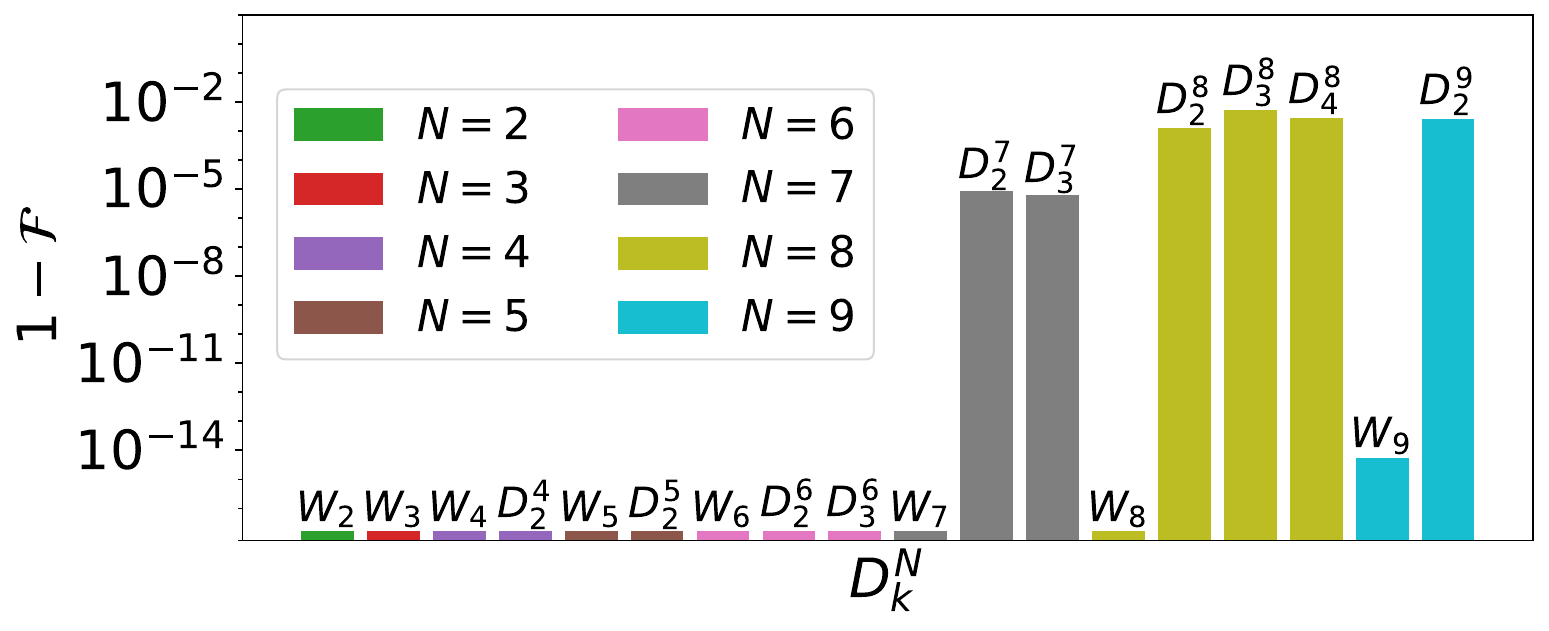}
    \caption{State preparation infidelity of $W$ and Dicke states within the NMR-like scheme, each one represented by a bar indicating the best result achieved. Different colors indicate different values of $N$.}
    \label{fig:nmr-fid-all}
\end{figure}

\begin{figure}
    \centering
    \includegraphics[width=0.95\columnwidth]{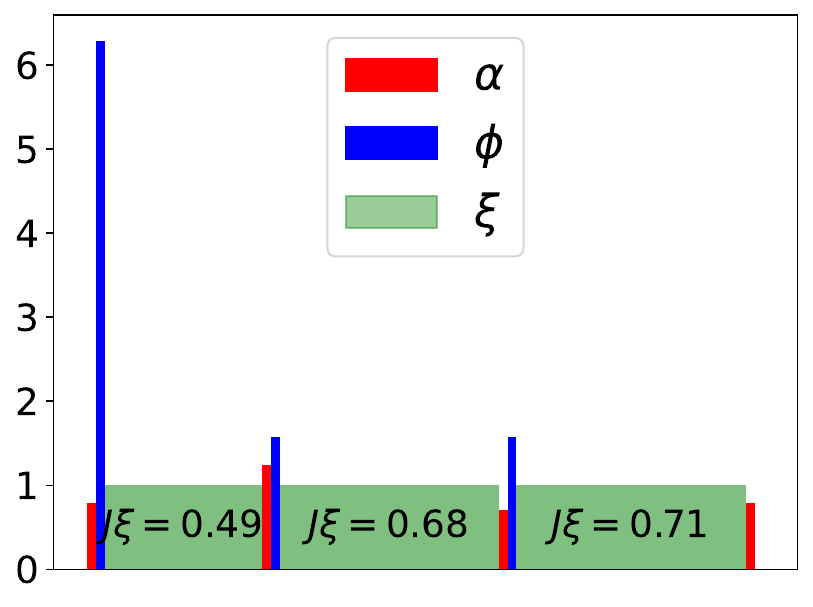}
    \caption{Example of optimal pulse for $\W5$ obtained within the NMR-like scheme.}
    \label{fig:nmr-pulse-all}
\end{figure}

In Fig.~\ref{fig:nmr-fid-all} we display the infidelity achieved for the preparation of $W$ and Dicke states of various size, showing that we obtained $1-{\cal F}<10^{-2}$ for each one of the reported cases.
As we can see, for low $N$, we managed to prepare them exactly up to numerical precision (i.e. we reached $1-{\cal F}\sim10^{-16}$), while for larger $N$ the optimization procedure is more and more challenging.
Furthermore, looking specifically at the case of Dicke states with $k>1$, we can see that the state preparation is harder not only when increasing $N$, but also when increasing $k$, and indeed we have not always been able to achieve exact preparation of states $\dicke{N}{k>1}$ even when their counterpart $\W{N}$ is prepared exactly.
As an example of optimal pulse, we display in Fig.~\ref{fig:nmr-pulse-all} one outcome of our optimization procedure for the preparation of $\W5$.

\begin{figure}
    \centering
    \includegraphics[width=0.95\columnwidth]{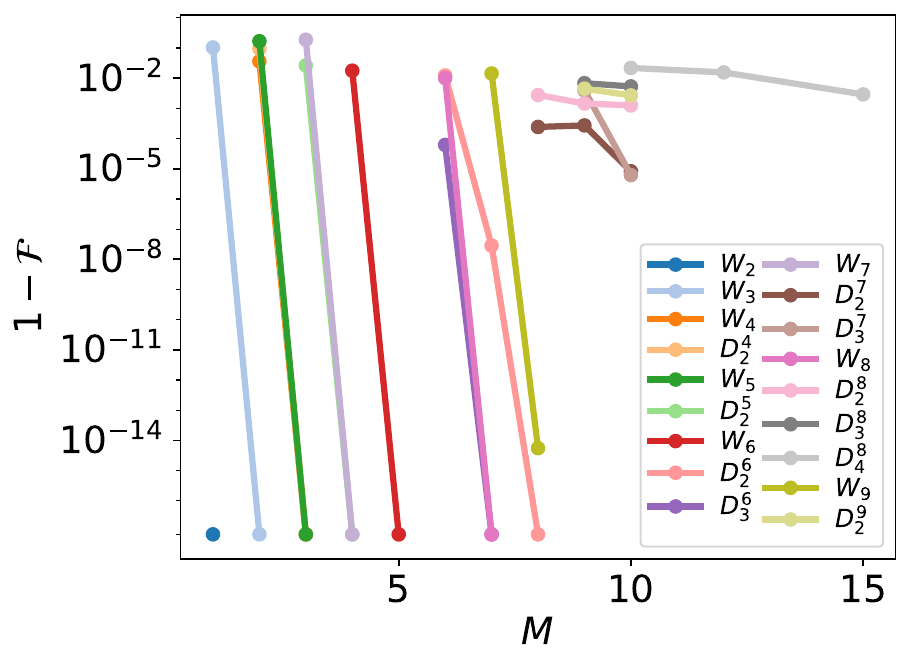}
    \caption{Best infidelity obtained as a function of $M$ within the NMR-like scheme.}
    \label{fig:nmr-fid-layers-all}
\end{figure}

In Fig.~\ref{fig:nmr-fid-layers-all} we display instead the fidelity change as a function of $M$ that we managed to obtain during our optimization procedure: notably, for low $N$, incrementing $M$ of even a single unity (i.e. adding a single layer $U_{xy}U_{ZZ}$ to the pulse) can bring the infidelity from $1-{\cal F}>10^{-2}$ to $1-{\cal F}\sim10^{-16}$, suggesting that there is a sharp cutoff in the minimal preparation time necessary for our target state within the current scheme.

\begin{figure}
    \centering
    \includegraphics[width=0.95\columnwidth]{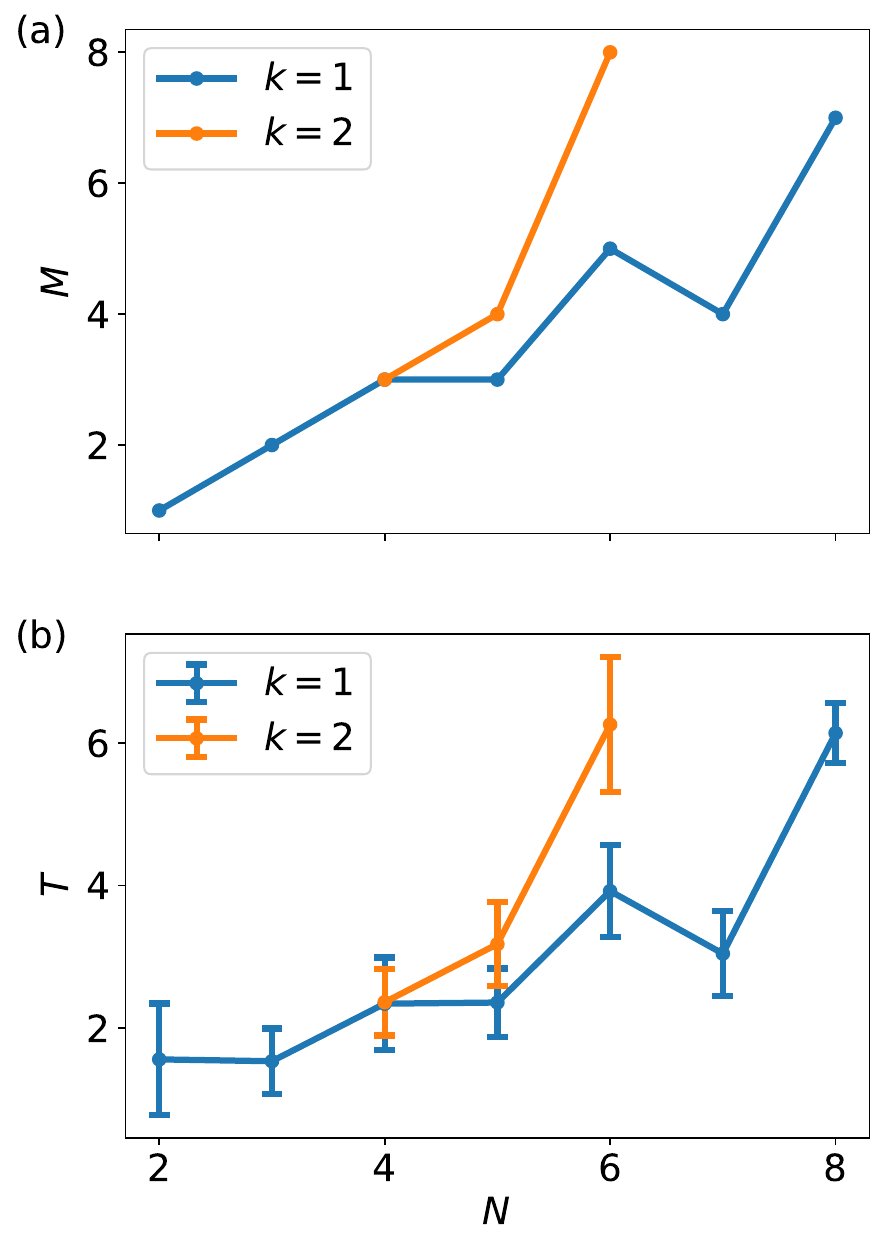}
    \caption{(a) Minimal number of interaction pulses $M$ necessary for the exact preparation of $W$ and Dicke (up to $k=2$) states, and corresponding (b) average preparation times (error bars indicate the standard deviation of the samples contributing to the average), within the NMR scheme.}
    \label{fig:nmr-nlayers-time-all}
\end{figure}

In Fig.~\ref{fig:nmr-nlayers-time-all} we investigate this more in depth, displaying the scaling of $M$ [Fig.~\ref{fig:nmr-nlayers-time-all}(a)] and of the total preparation time [Fig.~\ref{fig:nmr-nlayers-time-all}(b), averaged over several realizations] with respect to $N$, for pulses exhibiting $1-\mathcal F\sim 10^{-16}$. Since in an NMR scheme the global rotations are considered as instantaneous, the total duration time of the protocol is given by the sum of the durations of the Ising-interaction pulses, namely $T=\sum_i\xi_i$. We remark that, since the $\xi_i$s are bounded by finite values due to the periodicity of $U_{ZZ}$, this quantity is linearly proportional to $M$.
Looking at the results of Fig.~\ref{fig:nmr-nlayers-time-all}(b), we can see that the optimal preparation time of $W$ states (as well as the minimal $M$) is linear in $N$, in accordance with previous results~\cite{Chen+:17}.
On the other side, preparation time for Dicke states with $k=2$ seems to be superlinear instead, which also in this case aligns with previous results about preparation of Dicke states in analog schemes~\cite{keating_arbitrary_2016,ivanov_creation_2013,johnsson_geometric_2020,bond_global_2025}.
Due to the limited number of data points, we cannot test any further hypothesis about the actual scaling of the preparation time of Dicke states with $k>1$.
For the same reason, the same analysis cannot be done with $k\geq3$, neither fixing $N$ and looking and the scaling behavior of $T$ with $k$. This results also from the fact that states with $k>N/2$ can be obtained by a global $\pi$-pulse on the $x$ acted on the Dicke state with $N-k$ excitations, which can be incorporated as the last $U_{xy,M}$, hence not changing the duration of the pulse sequence. In other words, the minimal preparation time of $\dicke{N}{k}$ is the same as the one of $\dicke{N}{N-k}$.
This means that collecting data for many different values of $k$ requires optimization for larger and larger values of $N$, which rapidly becomes computationally impractical.

With the abundance of realizations collected using this approach, and given the achieved exact preparation of many of the target states (up to numerical precision), we perform also some statistical analysis on the optimal parameters obtained.
The idea behind this approach is to investigate the following question: are there correlations among the optimal parameters of $\W N$ and $\dicke Nk$ that can be exploited (for instance, for reducing the dimensionality of the pulses' parametrization)?
To this purpose, we restricted the analyzed results only to the ones exhibiting $1-{\cal F}\sim10^{-16}$, and we report here some considerations arising from simple analysis procedures.

\begin{figure}
    \centering
    \includegraphics[width=0.95\columnwidth]{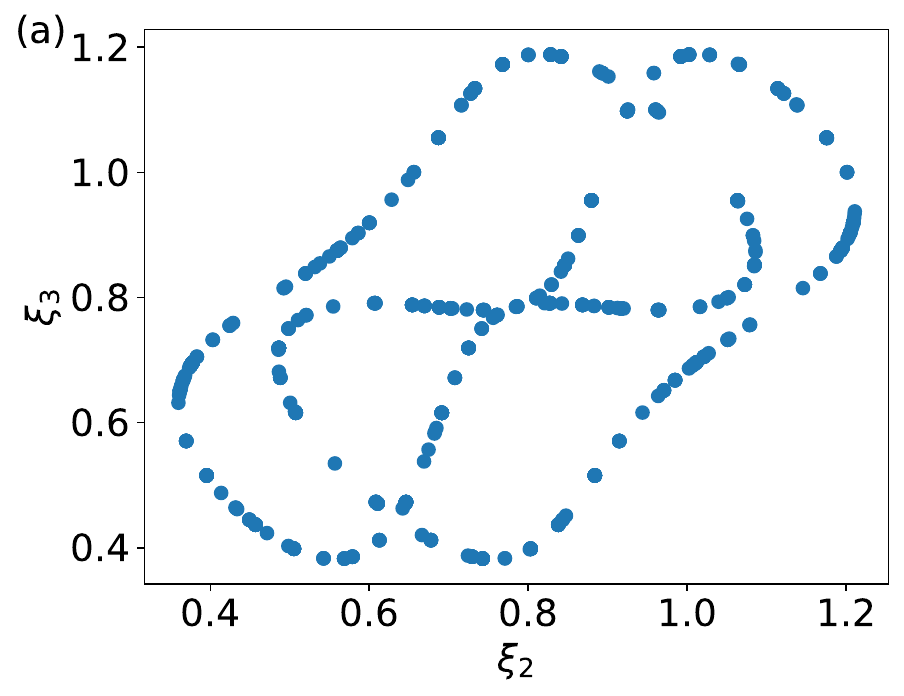}
    \includegraphics[width=0.95\columnwidth]{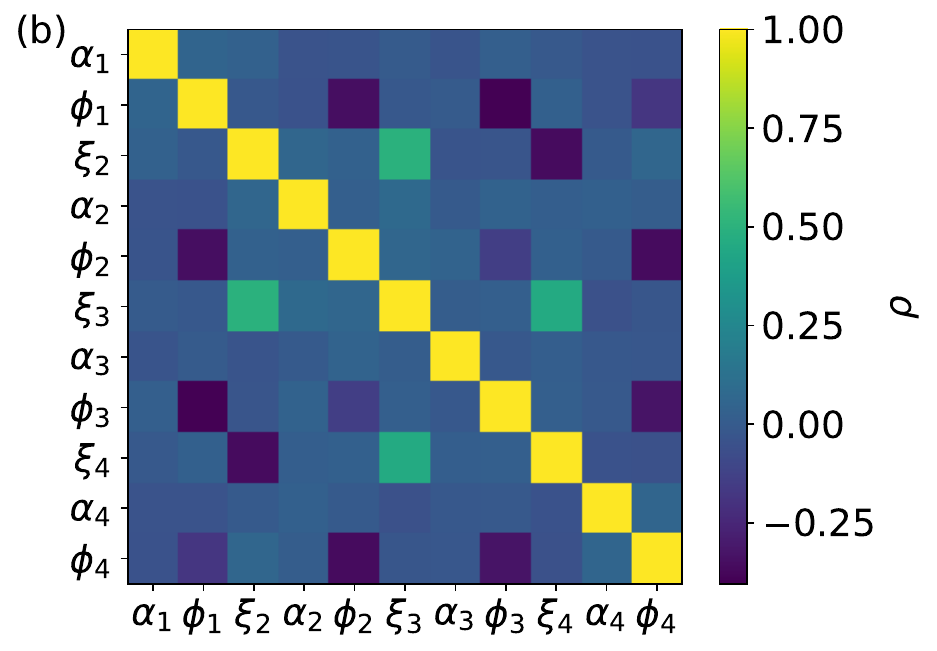}
    \caption{Parameters' correlations of optimal pulses for $\W5$ obtained within the NMR-like scheme (many realizations with $1-{\cal F}\sim10^{-16}$).
    (a) Scatter plot of the optimal values of a pair of parameters.
    (b) Pearson correlation coefficients $\rho$ for each pair of optimal parameters.}
    \label{fig:nmr-corr-all}
\end{figure}

In Fig.~\ref{fig:nmr-corr-all}(a) we display the scatter plot of optimal values of a pair of parameters. From this figure, one can promptly see that there is an evident correlation among the values assumed by such parameters, and many more correlations can be found considering other pairs. However, the functional forms of such correlations appear to be so cumbersome that the deep exploitation of such information seems impractical.
Looking at the distributions of the optimal parameters, we notice that some regions are totally avoided, suggesting a kind of higher invariance.
As a reference, we display also the Pearson correlation matrix~\cite{taylor_errors_1997} for every pair of parameters in the preparation of $\W5$ [Fig.~\ref{fig:nmr-corr-all}(b)], highlighting that, at least from a linear point of view, some pairs can be in general considered correlated while others anticorrelated (but often the correlation is actually highly nonlinear).

\begin{figure}
    \centering
    \includegraphics[width=0.95\columnwidth]{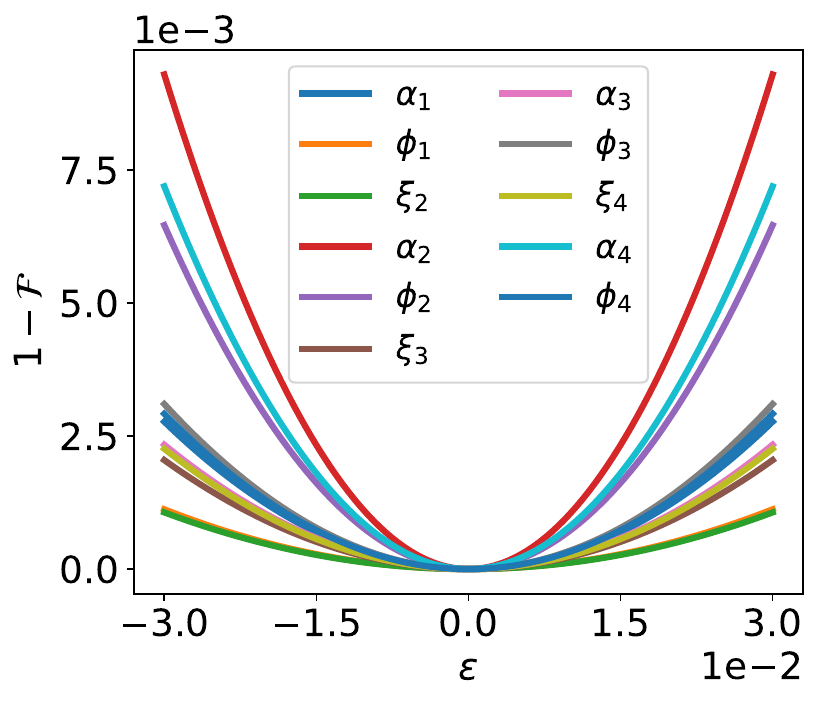}
    \caption{Infidelity change in the preparation of $\W5$ with an optimal NMR-like pulse when error $\varepsilon$ is introduced on one parameter at a time.}
    \label{fig:nmr-robustness1d-all}
\end{figure}

\begin{figure}
    \centering
    \includegraphics[width=0.95\columnwidth]{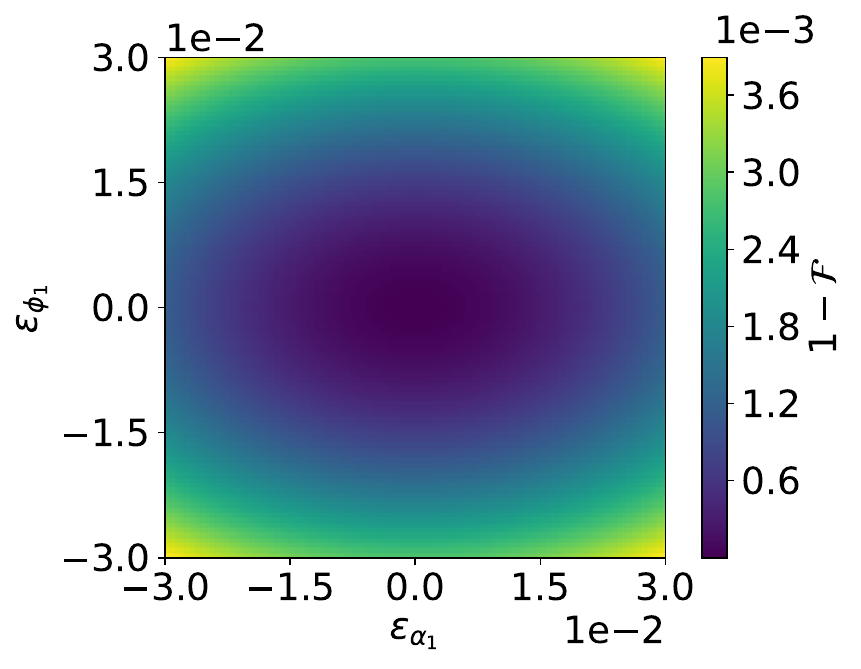}
    \caption{Infidelity change in the preparation of $\W5$ with an optimal NMR-like pulse when errors $\varepsilon$ are introduced on two parameters at the same time. In the figure, the effect of errors on $\alpha_1$ and $\phi_1$ is displayed.}
    \label{fig:nmr-robustness2d-all}
\end{figure}

We finally conclude this section investigating the potential robustness of the optimal pulses obtained. To this purpose, we look at the change of infidelity when introducing errors in the parameters. Following common error analysis in NMR state preparation~\cite{Vandersypen+Chuang:05}, we consider relative errors in the fields' intensities $\alpha_i$ and in the durations of the interaction pulses $\xi_i$, and absolute errors in the phases $\phi_i$, both positive and negative. In formulae, we are considering perturbations of the optimal parameters of the kind
\begin{equation}
    \alpha_j=\bar\alpha_j(1+\varepsilon_{\alpha_j}),\quad
    \phi_j=\bar\phi_j+\varepsilon_{\phi_j},\quad
    \xi_j=\bar\xi_j(1+\varepsilon_{\xi_j})
\end{equation}
where the overbar denotes the value of the optimal parameters.
An example of this kind of analysis can be found in Fig.~\ref{fig:nmr-robustness1d-all}, in which we addressed the preparation of $\W5$ when error on a single parameter occurs. As we can see, even in presence of single-parameter errors of reasonable magnitude (in the few percent range), the preparation infidelity is still $1-{\cal F}<10^{-3}$ for any of them.
Motivated by these results, we investigated also the infidelity behavior in case of two errors happening at the same time. Results are displayed in Fig.~\ref{fig:nmr-robustness2d-all} for a given pair of parameters, and highlight that, even in occurrence of two errors at the same time, it is still possible to achieve a nice state preparation with $1-{\cal F}<10^{-3}$.
We remark that, of course, not each one of the obtained optimal pulses has the same robustness against this kind of noise. In our case, we choose among the many realizations the one minimizing the overall sum of errors, namely the sum of the integrals of the curves displayed in Fig.~\ref{fig:nmr-robustness1d-all}. In principle, one can also directly include some form of robustness-aware term in the loss function already in the process of optimization in order to further improve on the flatness of the landscape reported here.

\section{$W$- and Dicke-state preparation with time-dependent pulses}\label{sec:tdep}
This section is devoted to the generation of Dicke and $W$ states using an alternative control approach, based on time-dependent control pulses. The basic aspects of this approach are first briefly presented 
in Sec.~\ref{sec:tdep-problem}. In Sec.~\ref{sec:tdep-results} we then discuss
the results obtained when using this approach 
for engineering $W$- and Dicke states.

\subsection{Continuous-wave control fields for state preparation and their optimization}\label{sec:tdep-problem}
We address here the case of preparation of quantum states with time-dependent control pulses. In this setting, we have electromagnetic fields continuously varying in time for driving the dynamics.
Unlike in the NMR case [cf. Sec.~\ref{sec:nmr-problem}], where the fields are just turned on and off with a given (strong) intensity, we have moreover that the fields' magnitudes are in general comparable with the interaction strength. As a result, we cannot split the dynamics in the alternation of global rotations and interactions, but it is necessary to take into account the whole Hamiltonian at each time.
The time-evolution operator of the system is in this case given by the 
most general form, which involves a 
time-ordered exponential, i.e.
\begin{equation}
    U(t)={\cal T}\exp\left\{-i\int_{t_i}^t H(t') {\rm d}t'\right\}
    \label{eq:tdep-U}
\end{equation}
where $t_{i}$ is the initial evolution time.

There are many recipes that one can use for addressing the optimization of pulses in this setting.
Among the most famous ones, the GRAPE (GRadient Ascent Pulse Engineering) algorithm~\cite{khaneja_optimal_2005}  discretizes the pulse in piece-wise constant steps and updates iteratively the height of each step for maximizing the preparation fidelity. The efficiency of the algorithm can be dramatically improved by including quasi-Newton \cite{de2011second} or Newton \cite{dalgaard2020hessian} updates. Methods that do not require a gradient and incorporate bandwidth limitation and smooth pulses' encoding in a few-parameters ansatz include CRAB (Chopped RAndom Basis)~\cite{caneva_chopped_2011}, encoding pulses in a chopped randomized Fourier basis, and the further dCRAB (dressed-CRAB)~\cite{rach_dressing_2015}, wisely stacking many CRAB iterations in order to avoid trapping in local minima of the landscape.
Moreover, other notable numerical optimization methods include the Krotov algorithm \cite{goerz2019krotov}, which can have enhanced convergence properties; GRAFS~\cite{lucarelli_quantum_2018} (GRadient Ascent in Function Space), putting particular attention on smooth controls; gradient methods to optimize over arbitrary control parametrizations \cite{motzoi2011optimal}, i.e.~especially where bandwidth limitation can be incorporated \cite{sorensen2020optimization, singh2023compensating}; as well as the very recent GEOPE~\cite{lewis_quantum_2025} (GEOdesic Pulse Engineering), exploiting the geometrical interpretation of unitary dynamics.

Here, taking inspiration from the previous literature, we choose a parametrization of our pulses in a truncated sine basis
\begin{equation}\label{eq:sine-pulse}
B_{x,y}(t)=b^{(x,y)}_0+\sum_{m=1}^Mb^{(x,y)}_m\sin(m\pi t/T) \:,
\end{equation}
where $T$ is the total duration of the pulse (in appropriate units) and $b_0$ takes into account the presence of a constant (DC) component.
Such parametrization, despite its simplicity, allows us to encode smooth (since the basis is a superposition of smooth functions) and bandwidth-limited (through appropriate choice of the cutoff $M$) pulses with just a few parameters; this facilitates potential further closed-loop optimization~\cite{motzoi2011optimal,sorensen2020optimization}.

As in the NMR scheme, the optimization is aimed at maximizing the target-state fidelity $1-{\cal F}_{t=t_f}$ [cf. 
Eq.~\eqref{eq:fidelity}] as the main figure of merit, together with some additional terms that account for constraints on global properties of the control fields (see further details in Appendix~\ref{app:CodeTimeDep}), and the optimization parameters are the magnitudes of each component of the sine-decomposition, separately for the $x$ and $y$ fields, denoted by $\{b_m^{(x)}, b_m^{(y)}\}$ in Eq.~\eqref{eq:sine-pulse}.

Unlike in the NMR scheme, here it is more computationally challenging -- due to the 
more demanding numerical evaluation of the time evolution -- to perform optimization also over the value of $M$ which is necessary to achieve the minimal fidelity threshold. For this reason, we propose an alternative approach for designing ``simplified'' pulses.
Starting from an optimized pulse achieving a fidelity above a set threshold (we set in our case $1-{\cal F}_{t=t_f}=10^{-2}$), we remove from the parametrization the ``least significant’’ component, i.e. the one with the smallest absolute magnitude, and rerun the gradient optimization with the reduced parametrization, using the remaining previous optimal parameters as the new initial seed. If the desired fidelity is once again obtained with the new parametrization, the procedure is repeated iteratively until the minimal fidelity is no longer achievable within a given number of attempts.
The final successful pulse is interpreted as the “simplest’’ pulse compatible with the target fidelity threshold, and we denote as $\tilde M$ the number of parameters left (for each pulse).
This follows the idea that these pulse shapes are easier to implement experimentally due to their easier functional form, and thus less prone to experience errors coming from hardware imperfections.

Due to the complexity of the optimization landscape, which does not allow starting optimization from totally random initial guesses, we initialized the optimization with the guess $B_{x,y}(t)=\sin(\pi t/T)$, i.e. a raise-and-fall of the fields, which is a standard choice in the preparation of entangled states with Rydberg atoms~\cite{in_guess}.
We extract from this seed the corresponding coefficients in our parametrization and generate multiple initial points by sampling normally distributed values around them. Optimization is then performed on the initial guesses exhibiting the lowest loss. This results in a restricted, “informed’’ version of the multistart strategy used in the NMR case [cf. Sec~\ref{sec:nmr-problem}].
Because of the computational intractability of performing too extensive simulations, we will not present here statistics on the parameters' values.

Further details about code implementation can be found in Appendix~\ref{app:CodeTimeDep}.

\subsection{Time-dependent control pulses for $W$- and Dicke-state
preparation: results and discussion}\label{sec:tdep-results}
We present here the results obtained for the preparation of $W$ states with time-dependent pulses with up to $N=8$ qubits.
Given the computational intractability of performing direct optimization on the value of $M$, we decided to set initially $M=9$, and to later reduce the number of parameters with the iterative simplification procedure described in Sec.~\ref{sec:tdep-problem}. The current Section will be thus divided accordingly in two parts.
Throughout this Section, we will indicate with just $\cal F$ the target-state fidelity ${\cal F}_{t=t_f}.$

Unlike in the NMR case, here we addressed the preparation of Dicke states with $k>N/2$ too. This choice comes from the fact that, within the proposed pulses' parametrization, it is not a priori trivial how to modify the pulse sequence in order to turn a state $\dicke{N}{k}$ into its $\dicke{N}{N-k}$ counterpart (while it was in the NMR scheme, cf. Sec.~\ref{sec:nmr-results}).
In principle, a global NOT gate can be achieved sending a square pulse of area $\pi$ and intensity $B_x\gg J$ after the end of the parameterized sequence. However, this will elongate the duration of the sequence and is not always guaranteed to be feasible in experiments of this kind; moreover, the overall time-envelope of $B_x$ will fall out of the functional forms achievable with the parametrization in Eq.~\eqref{eq:sine-pulse}.

All the pulses are evolved for a total (dimensionless, in units of $J^{-1}$) time of $T=4$.

\begin{figure}
    \centering
    \includegraphics[width=0.95\columnwidth]{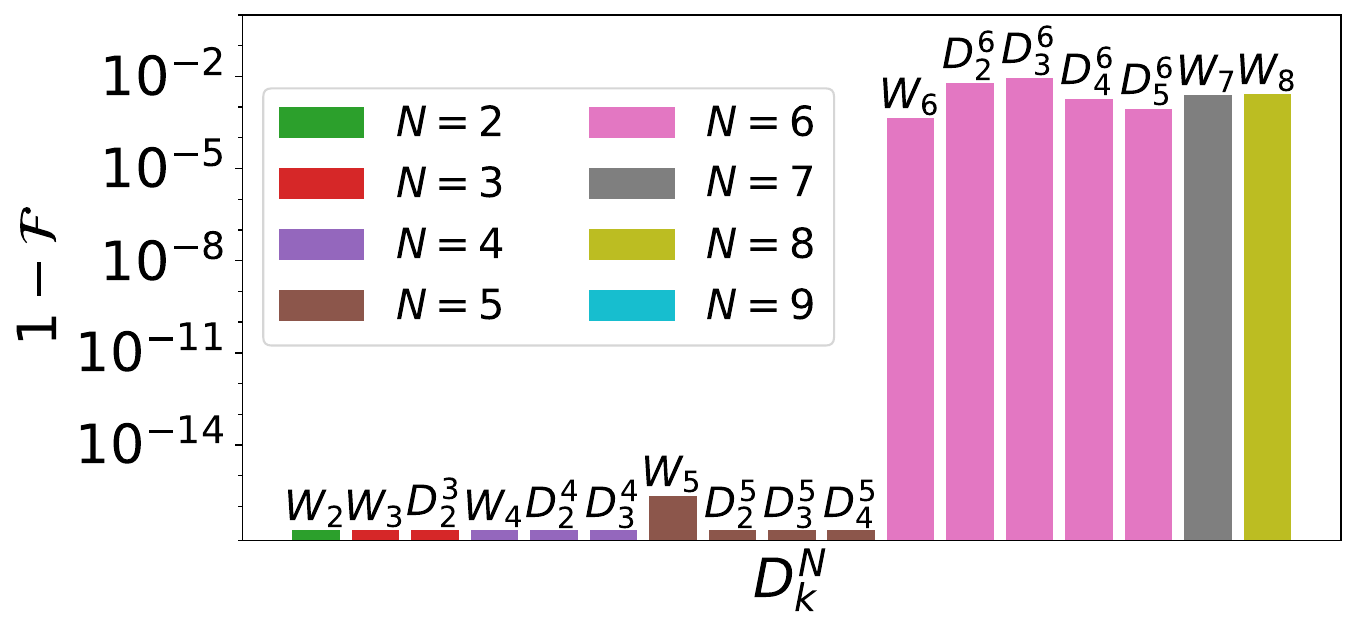}
    \caption{State preparation infidelity of $W$ and Dicke states within the time-dependent pulses scheme, each one represented by a bar indicating the best result achieved. Different colors indicate different values of $N$.}
    \label{fig:tdep-fid-all}
\end{figure}

\begin{figure}
    \centering
    \includegraphics[width=0.95\columnwidth]{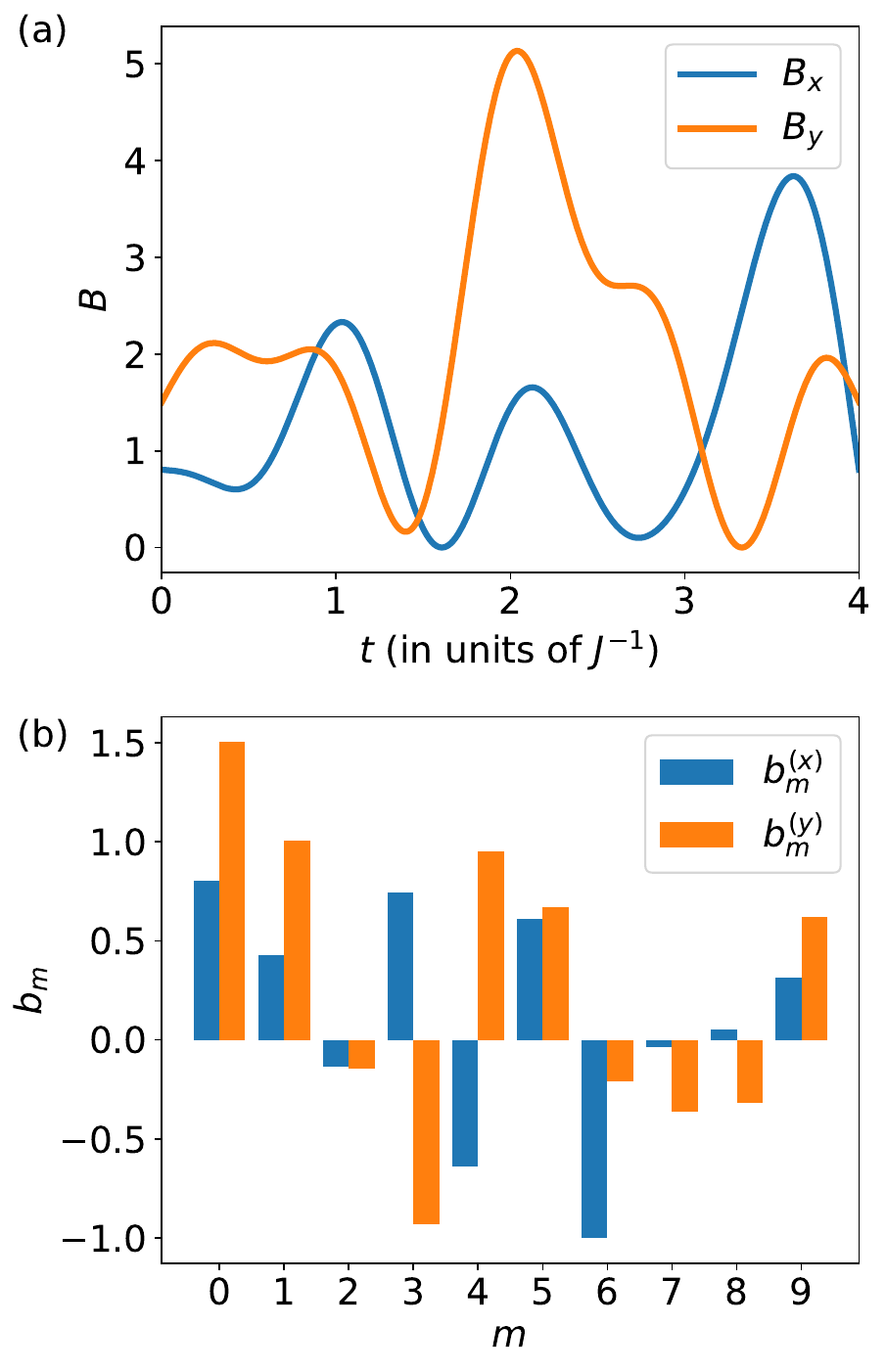}
    \caption{Example of optimal pulse for $\W4$ obtained within the time-dependent pulses scheme.
    (a) Pulse shapes of the fields in the time domain.
    (b) Magnitude of the parameters used for building the pulse shapes above.}
    \label{fig:tdep-pulse-all}
\end{figure}

\paragraph{Fixed $M$ optimization}
In Fig.~\ref{fig:tdep-fid-all} we display the infidelity achieved for the preparation of $W$ and Dicke states of various size, showing that we managed to prepare them always with $1-{\cal F}<10^{-2}$. As expected, when increasing the number of atoms $N$ in our system, the optimization becomes more and more challenging, hence the infidelity increases as well. The same behavior is not evident as a function of $k$, potentially also because of the few data points collected.
We observe then an abrupt fidelity jump at $N = 6$: this can be likely attributed to the fixed number of optimization parameters. Indeed, similarly to the NMR case, where adding a single further batch of parameters leads to improvements of several orders of magnitude of infidelity, here the use of too few components becomes restrictive once a certain system size is exceeded [cf. Fig.~\ref{fig:nmr-fid-all} from Sec.~\ref{sec:nmr-results}].
As an example of optimal pulse, we display in Fig.~\ref{fig:tdep-pulse-all} one outcome of our optimization procedure for the preparation of $\W4$.

A study of the minimal preparation time analogous to the NMR case [cf. Sec.~\ref{sec:nmr-results}] is in this case more challenging, especially due to the dependence of $T$ on many parameters. Indeed, one might think to just increase $J$ and thus reduce the value of $T$ in our protocol, however this could easily clash against the feasibility of the proposed scheme. Indeed, in a neutral atom experiment, $J$ is related to the distance among atoms: even if the decay is very fast (proportional to the inverse sixth power of the distance), getting atoms too close to each other can introduce non-negligible next-nearest-neighbors (or even longer range) interactions, which would make us fall into a different system from the one we are studying here.
Moreover, even including the possibility to increase $J$ to high values (this would be possible without introducing spurious interactions, for instance, if $N$ is quite large), the resulting pulse would exhibit fast large oscillations that are unlikely to be performed due to both limited bandwidth and limited max power of the hardware. In other words, in order to reduce $T$, in addition to finding an optimal value of $J$, one should in principle also resort to pulses with a lower $M$, and finding the right compromise balancing all of these parameters is computationally challenging. 
We refer to Sec.~\ref{sec:implementRydberg} for more details about experimental implications.

\begin{figure}
    \centering
    \includegraphics[width=0.95\columnwidth]{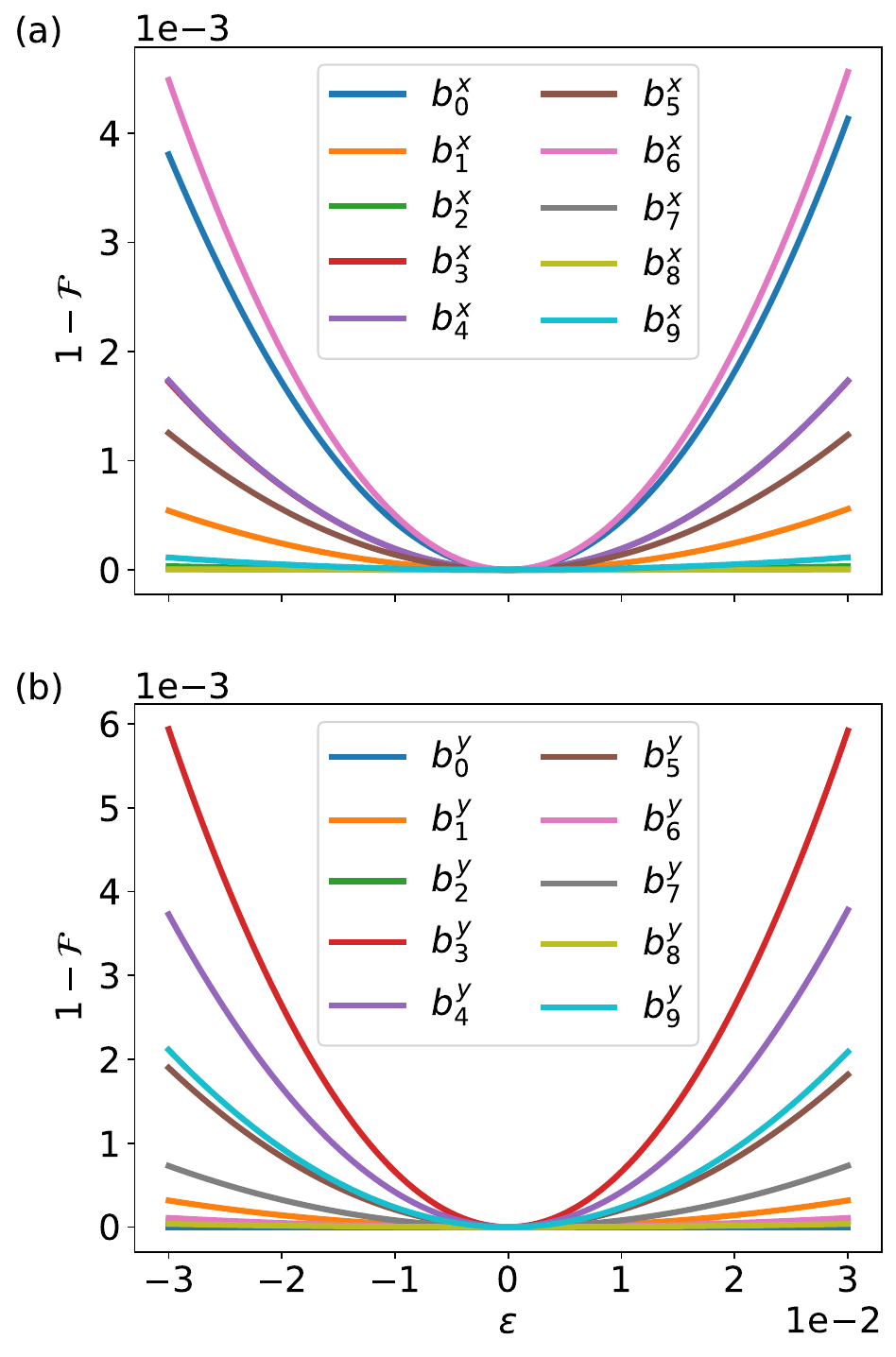}
    \caption{
    Infidelity change in the preparation of $\W4$ with an optimal time-dependent pulse when error $\varepsilon$ is introduced on one parameter at a time.
    (a) Errors on the parameters of the field $B_x$.
    (b) Errors on the parameters of the field $B_y$.
    }
    \label{fig:tdep-robustness1d-all}
\end{figure}

\begin{figure}
    \centering
    \includegraphics[width=0.95\columnwidth]{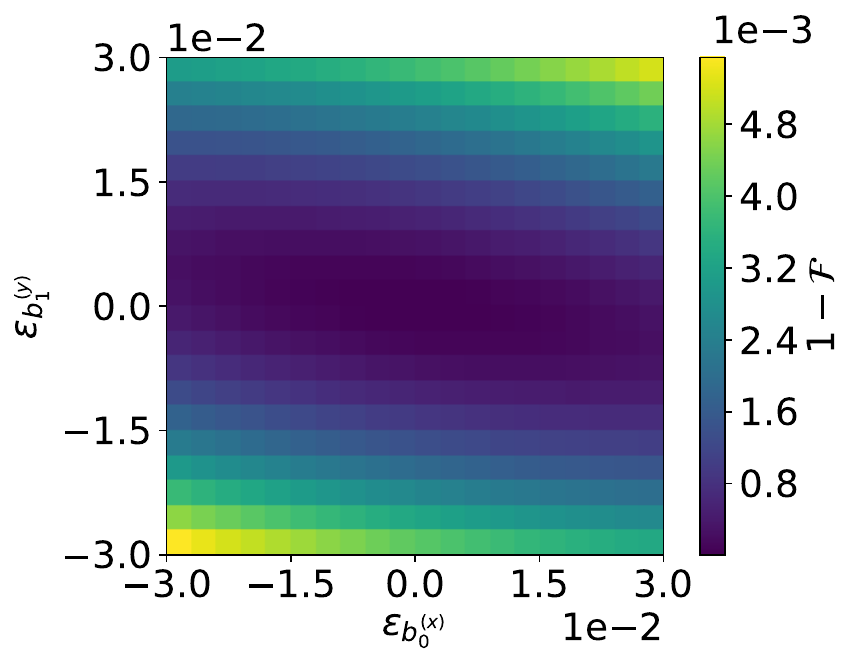}
    \caption{Infidelity change in the preparation of $\W4$ with an optimal time-dependent pulse when errors $\varepsilon$ are introduced on two parameters at the same time. In the figure, the effect of errors on $b_0^{(x)}$ and $b_1^{(y)}$ is displayed.}
    \label{fig:tdep-robustness2d-all}
\end{figure}

We address finally the problem of the robustness of the pulses. In this case we consider percentage error on each one of the optimal parameters, both positive and negative, namely we consider perturbations
\begin{equation}
b_m^{(x,y)}=\bar b_m^{(x,y)}[1+\varepsilon_{b_m^{(x,y)}}]
\qquad (\:m=0,\ldots, M\:) \:,
\end{equation}
where the overbar denotes the optimal parameter values.
An example of this kind of analysis can be found in Fig.~\ref{fig:tdep-robustness1d-all}, in which we addressed the preparation of $\W4$ when error on a single parameter occurs. As we can see, even in presence of single-parameter errors of reasonable magnitude, the preparation infidelity is still $1-{\cal F}<10^{-2}$ for any of them.
Motivated by these results, we investigated also the infidelity behavior in case of two errors happening at the same time. Results are displayed in Fig.~\ref{fig:tdep-robustness2d-all} for a given pair of parameters, and highlight that even in occurrence of two errors at the same time, it is still possible to achieve a nice state preparation with $1-{\cal F}<10^{-2}$.
Just like in the NMR case, we remark also here that not all of the obtained optimal pulses have the same robustness against this kind of noise. For this reason, we choose among the many realizations the one minimizing the overall sum of errors, namely the sum of the integrals of the curves displayed in Fig.~\ref{fig:tdep-robustness1d-all}. Also for the present setting, the optimization problem can be promptly generalized to include a robustness-aware term in the loss function. A notable example of this kind of approach is reported in~\cite{deLima+:25} for the all-to-all interaction case, in which the authors leveraged the low-dimensional space of the fully permutational-symmetric system to carry out computationally heavy robustness-oriented optimization.

\paragraph{Iterative components' removal}
After having presented the results for the optimization of time-dependent pulses with a fixed number of parameters, we show here the outcomes of our iterative procedure for removing components from the optimal pulses described in Sec.~\ref{sec:tdep-problem}.

\begin{figure}
    \centering
    \includegraphics[width=0.95\columnwidth]{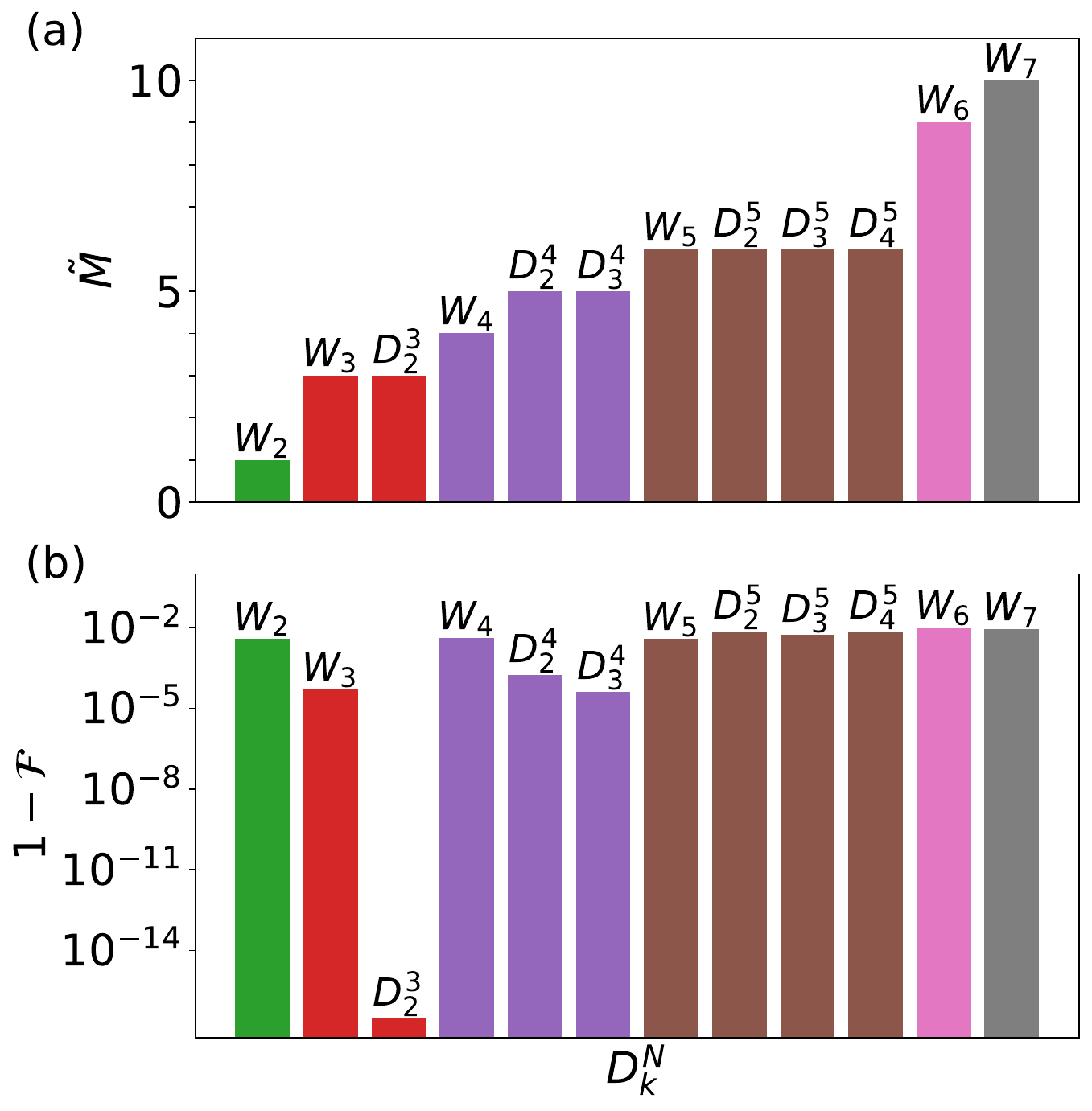}
    \caption{(a) Minimal number of components $\tilde M$ and (b) corresponding state preparation infidelity achieved for the preparation of $W$ and Dicke states within the time-dependent pulses scheme after that the iterative components' removal procedure has been applied, with a minimal fidelity threshold of $\mathcal F=0.99$. Each state is represented by a bar indicating the best result achieved. Different colors indicate different values of $N$.}
    \label{fig:tdep-rem-nharm-all}
\end{figure}

In Fig.~\ref{fig:tdep-rem-nharm-all} we show the infidelity achieved for many $W$ and Dicke states at the end of our procedure and the minimal number of components $\tilde M$ necessary to exceed the minimal fidelity threshold (in our case $0.99$).
As we can see, states with lower $N$ require only few components, whereas larger states require a progressively larger $\tilde M$, in qualitative analogy with the NMR-based approach and in accordance with general expectations. The infidelity also increases with $N$, reflecting the growing difficulty of accurately preparing more complex entangled states.
On the other side, we see, as in the previous subsection, that the value $\tilde M$ does not seem to be affected by increasing $k$ in the range explored.
This makes us conjecture that the time-dependent scheme is less sensitive to the Hamming weight of the target state, when compared to the NMR one.

\begin{figure}
    \centering
    \includegraphics[width=0.95\columnwidth]{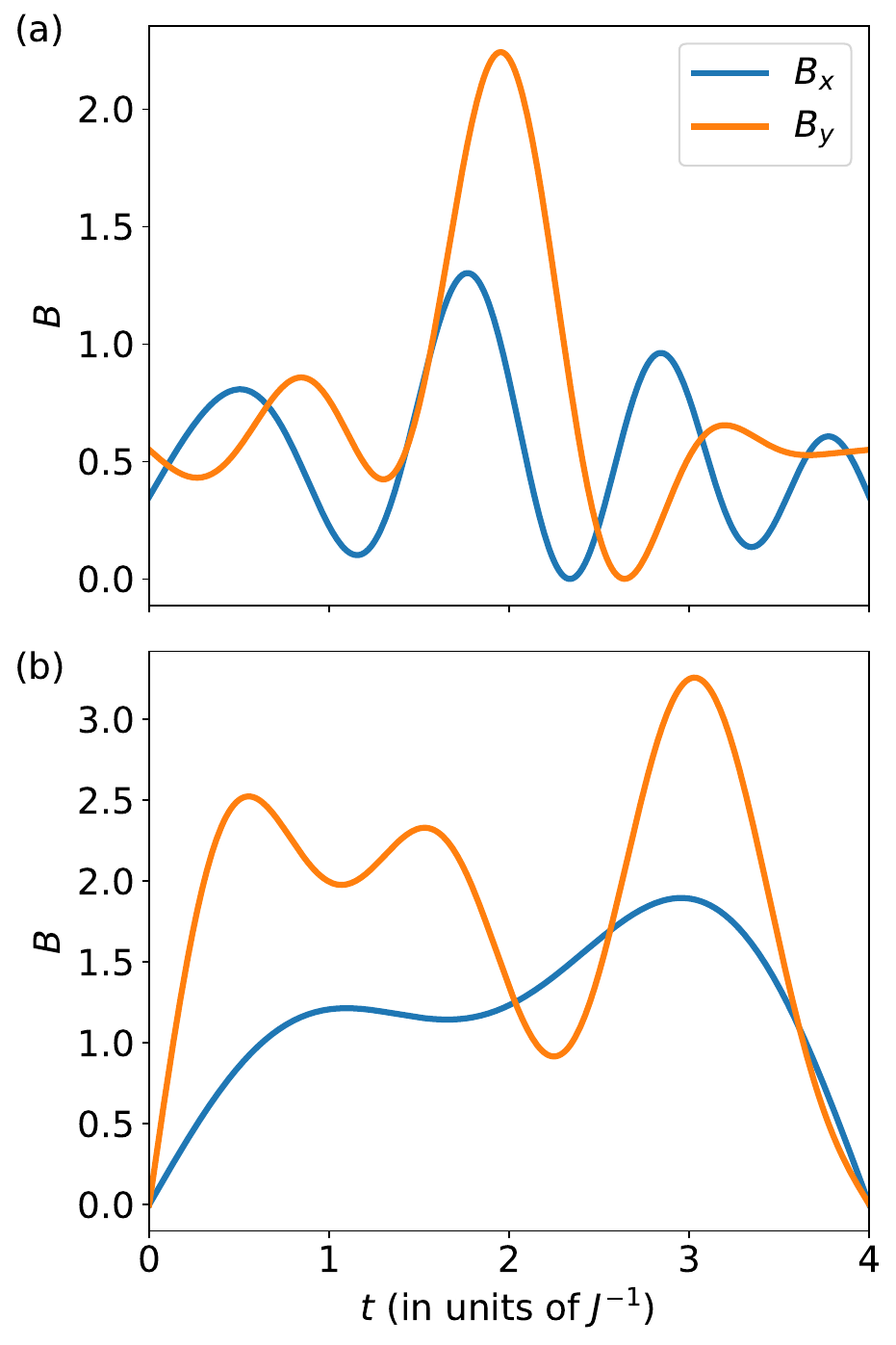}
    \caption{Example of optimal pulse for $\dicke32$ [cf. Fig.~\ref{fig:tdep-rem-components-all} for parameters' values] obtained within the time-dependent pulses scheme (a) before and (b) after that the iterative components' removal procedure has been applied, with a minimal fidelity threshold of $\mathcal F=0.99$.}
    \label{fig:tdep-rem-pulse-all}
\end{figure}

\begin{figure}
    \centering
    \includegraphics[width=0.95\columnwidth]{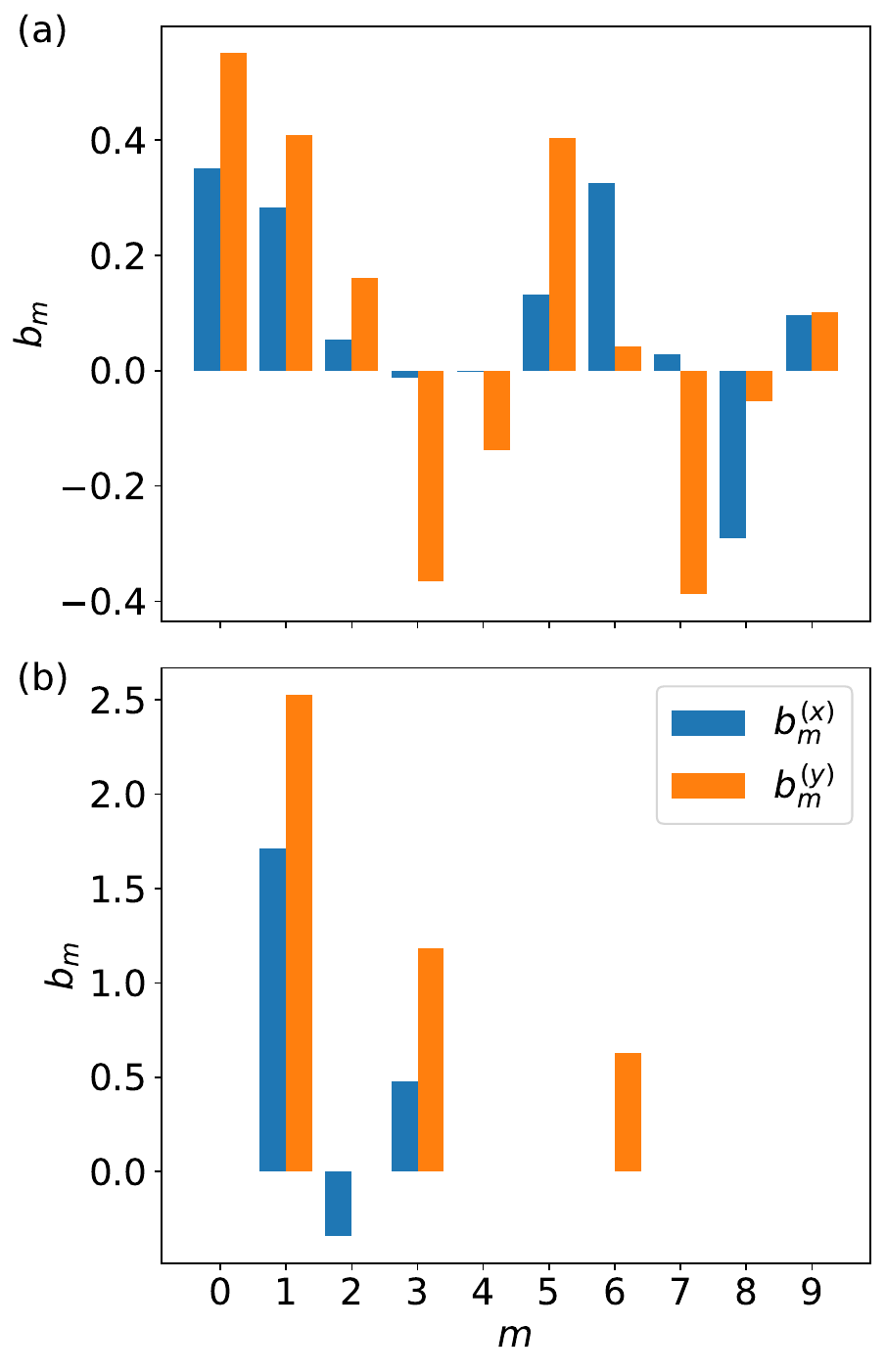}
    \caption{Example of optimal pulse parameters for $\dicke32$ [cf. Fig.~\ref{fig:tdep-rem-pulse-all} for pulse shapes in time domain] obtained within the time-dependent pulses scheme (a) before and (b) after that the iterative components' removal procedure has been applied, with a minimal fidelity threshold of $\mathcal F=0.99$.}
    \label{fig:tdep-rem-components-all}
\end{figure}

In Fig.~\ref{fig:tdep-rem-pulse-all}, we display the pulses obtained for the preparation of $\dicke32$ before and after iterative components' removal, together with their respective parameters' values in Fig.~\ref{fig:tdep-rem-components-all}.
Comparing the two results, the new pulse is visibly smoother and exhibits significantly fewer oscillations. Notably, we also remark that with this procedure we have been able to recover the well-known single-frequency solution for preparing the state $\dicke21$ using only the $k=1$ component, illustrating its effectiveness.

\begin{figure}
    \centering
    \includegraphics[width=0.95\columnwidth]{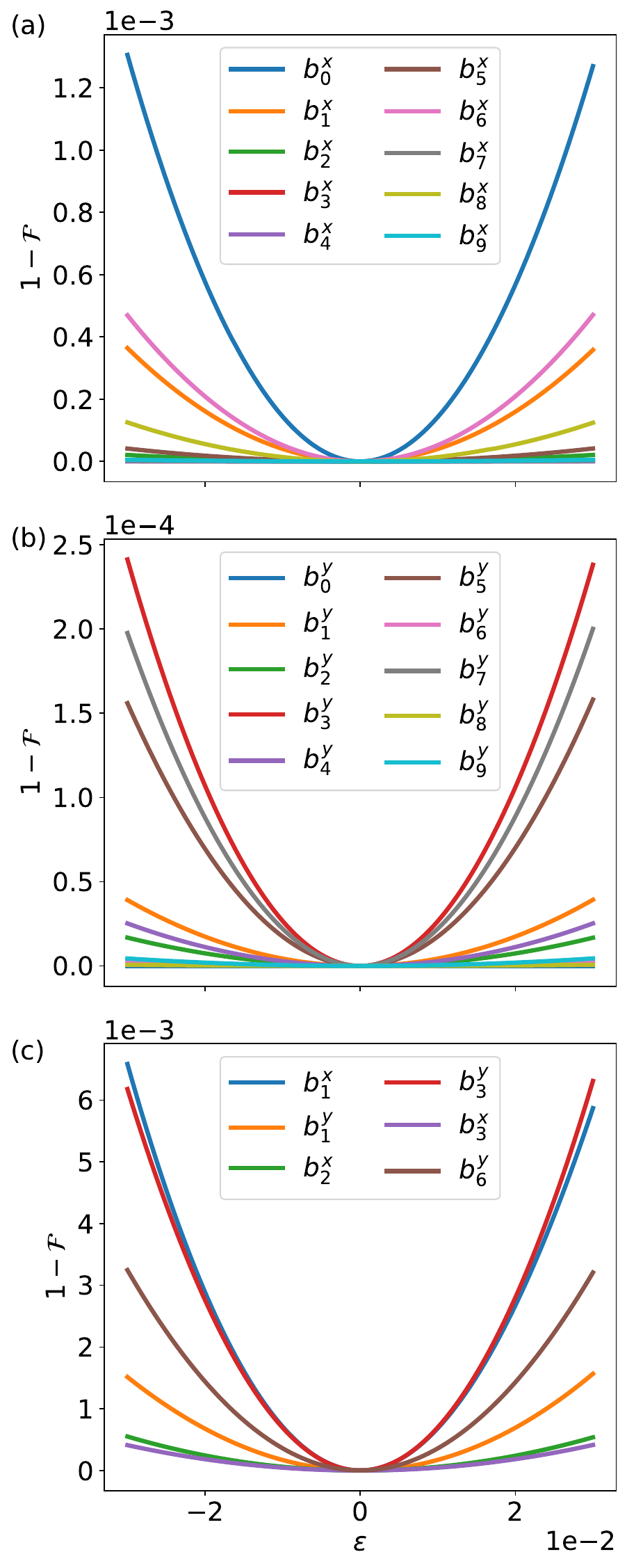}
    \caption{Infidelity change in the preparation of $\dicke32$ with an optimal time-dependent pulse when error $\varepsilon$ is introduced on one parameter at a time. Comparison between (a,b) before and (c) after that the iterative components’ removal procedure has been applied, with a minimal fidelity threshold of $\mathcal F=0.99$.}
    \label{fig:tdep-rem-robustness-all}
\end{figure}

We pursue the same kind of analysis about robustness, adding percentage errors on each of the parameters.
We can see that also in the case of simplified pulses the robustness can be achieved as well, but the path to it is once again not straightforward.
In particular, we applied the following procedure. We first started from an initial pulse with a given fixed $M$, and we applied the iterative components' removal procedure. Once reached the minimum value of components $\tilde M$, we produced several optimized outcomes using the found parametrization, and then selected the most robust one against noise. Also in this case, our robustness criterium has been the sum of integrals of the single-parameter error curves.
The investigation of a noise-aware optimization for this last case (instead of just producing many outcomes) is of interest.
Moreover, tradeoffs between minimal number of components $\tilde M$ and robustness can be investigated as well.
Results obtained in this way are visible in Fig.~\ref{fig:tdep-rem-robustness-all}.

\section{Implications for experiments with neutral atoms in optical tweezers} \label{sec:implementRydberg}

In what follows we discuss the feasibility of realizing the 
state-preparation protocols proposed in Secs.~\ref{sec:nmr} and \ref{sec:tdep} above
in realistic neutral-atom systems.

The platform from which we took inspiration for the current work is a neutral atoms quantum simulator~\cite{scholl_quantum_2021}.
In this kind of platforms, up to hundreds of atoms can be trapped and confined in a region of few micrometers of diameter.
Current technology, based on optical tweezers, opens up the possibility to move around atoms in a 2D plane, allowing to arrange them in an 
almost arbitrary topology~\cite{Schymik+:20}.
The roles of the logical $\ket0$  and 
$\ket1$ states of a qubit are played by the electronic ground state $\ket g$ and a highly-excited Rydberg state $\ket r$ of an atom, respectively. Two atoms 
in Rydberg states corresponding to the same principal quantum number 
interact through van-der-Waals interaction 
\begin{equation}
\begin{split}
H_{\textrm{vdW}} &=\sum_{n<n'}\frac{C_6}{R^6_{nn'}}\ket{r_n r_{n'}}\bra{r_n r_{n'}}\\
&=\sum_{n<n'}\frac{C_6}{4R^6_{nn'}}\:(\sigma_n^z\sigma_{n'}^z+\sigma_n^z+\sigma_{n'}^z+\mathbbm1) \:,
\end{split}
\end{equation}
where $R_{nn'}$ is the distance between the atoms $n$ and $n'$, with $C_6$ being the 
van-der-Waals interaction constant.

As one can see, the interaction term enables  to perform multi-qubit gates (and, in particular, controlled operations~\cite{Browaeys+Lahaye:20}), and its dependence on the inverse sixth power of the interatomic distance makes it very strong for nearby atoms as well very fast decaying once distance is increased.
In particular, one can define two regimes in Rydberg-atoms experiments, depending on the relative strength of the interaction and the resonant field $B_x$ coupling $\ket g$ to $\ket r$. In case of $B_x\ll C_6/R_{nn'}^6$, we talk about the Rydberg blockade regime~\cite{barredo_demonstration_2014}, in which the interaction is dominant and the 
simultaneous (resonant) population of two (interacting) Rydberg levels is practically forbidden. On the other hand, we have the antiblockade regime~\cite{ates_antiblockade_2007}, in which interacting Rydberg levels can become significantly populated at the same time instead.
The boundary between these two conditions is defined at the equality $B_x=C_6/R_{nn'}^6$. It is easy to note that this condition defines a precise distance $R_B=C_6/B_x$, the so-called Rydberg-blockade radius, which determines the spatial separation needed between atoms in order to sit in one of the two regimes.

Concerning the results of the present work, we want to prepare a 1D chain of qubits with periodic boundary conditions. This can be easily achieved placing the atoms equally spaced on a ring. The interaction coefficient will then be $J=C_6/R^6$, $R$ here denoting the (constant) atoms spacing.
We moreover considered only nearest neighbors interactions, meaning that we need to be in the approximation of longer-range interaction much lower than $J$ and the field strengths: this is indeed ensured by the fast spatial decay $\propto R^{-6}$ of the interaction strength.

Furthermore, we underline that we want to prepare not only $W$ states, but also Dicke states. Even if sitting in the blockade regime would be ideal for the preparation of $W$ states (indeed, the all-to-all blockade is the most natural setting for it), this would be detrimental for the preparation of Dicke states, which in general contain neighboring excitations. For this reason, we will consider a system setting where the relation $B_x\gtrsim J$ holds, where we have both non-negligible effect of the interactions and at the same time the possibility to prepare neighboring excitations (the actual choice of maximum ratio $B_x^{\rm max}/J$ depends on using either the NMR-like or time-dependent pulses scheme).

For instance, a Rydberg atom made out of Rubidium with principal quantum number of the Rydberg level $n_{\rm Ryd}=60$ has an interaction coefficient $C_6=8.66\times10^5\ {\rm rad}\ \mu\rm m^6/\mu\rm s$, meaning that for an interatomic distance of $R=6.65\ \mu\rm m$ the interaction strength is $J\approx10\ {\rm rad}/\mu\rm s$.
Current commercially available quantum simulators, using the Rydberg atoms just described and belonging to the case of time-dependent pulses, have the possibility to reach $B_x^{\rm max}=12.5\ {\rm rad}/\mu{\rm s}$~\cite{rava_benchmarking_2025}, which is in very nice agreement with the range of parameters we are aiming at.
We underline, moreover, that such machines also have limitations on the minimal interatomic distance $R_{\rm min}=5\ \mu{\rm m}$~\cite{rava_benchmarking_2025}, which has further motivated the present study of preparation of multiqubit entangled states with only nearest-neighbors interaction instead of all-to-all.

We finally provide some considerations about the preparation times of the symmetric states targeted in this work, analyzing separately the NMR scheme and the time-dependent pulses one.
It is known that the lifetime of a Rydberg level in neutral atoms experiments is of the order of $100\ \mu s$, hence it is desirable to design preparation protocols not exceeding neither approaching this duration.

In the NMR scheme, we are preparing the states through a pulse sequence consisting in the alternation of global rotations, performed by strong delta-shaped magnetic fields, and Ising-interaction pulses. Given the (approximately) instantaneous nature of the rotation pulses, we can define the total duration of our protocol as the sum of durations of the interactions pulses, i.e. $T=\sum_i\xi_i$. We recall that the periodicity of the interaction operator $U_{ZZ}(\xi_j)=e^{-i\xi_j H_{ZZ}}$ is $\pi$ ($\pi/2$ if $N\geq3$), and that $\xi_j$ is expressing a time in units of the inverse interaction strength $J^{-1}$, namely $\xi_j=Jt_j^{\rm exp}$, where $t_j^{\rm exp}$ denotes the actual experimental time of the $j$-th interaction pulse. 
For these reasons, the maximal experimental time is upper bounded by $T_{\rm exp}\leq T^{\rm max}_{\rm exp}=M\pi/J$ [$M\pi/(2J)$ if $N\geq3$], where $M$ counts the number of interaction pulses.
Given a reasonable value of $J=10\ {\rm rad}/\mu\rm s$ (as previously discussed), this means that our preparation scheme can be performed in less than a microsecond, which is two orders of magnitude shorter than the 
typical radiative lifetime of Rydberg states.

We move then to the time-dependent pulses scheme. We recall that also in this case the simulations have been performed in a timescale of units of $J^{-1}$, hence, being the simulation time fixed at $T=4$, the actual  duration of the experiment is simply given by $T_{\rm exp}=4\:J^{-1}$.
Also in this case, for a value of $J=10\ {\rm rad}/\mu\rm s$, we could conclude that the experiments can be performed in less than a microsecond.
However, we must highlight that in this context we can encounter limited-bandwidth issues, i.e. constraints given by the fact that pulses cannot change their intensity too fast.
Even if we addressed this issue by limiting the maximum frequency of the pulses as well the maximum value of fields, this problem can be further tackled by slowing down the dynamics through a lower interaction strength $J$. Even in this latter case, this keeps us still far away from the Rydberg state lifetime: for instance, using $J=1\ {\rm rad}/\mu\rm s$ makes the pulses last only $T_{\rm exp}=4\ \mu\rm s$, which is much lower the $100\ \mu \rm s$ coherence time of a Rydberg state.
In addition, our method does not impose a priori any constraint on duration and bandwidth of the pulses, hence it can be easily customized to specific capabilities and needs.

\section{Summary and Conclusions} \label{sec:SumConcl}

 To summarize, in this paper we investigated -- using quantum optimal-control techniques -- the preparation of $W$ and Dicke states in a ring-shaped array of qubits with nearest-neighbor Ising-type ($zz$) coupling and transverse ($x$ and $y$) global control. This model study was primarily motivated by the compelling need to design fast and robust schemes for engineering highly-entangled multiqubit states in ensembles of neutral atoms in optical tweezers; more specifically, this work is of direct relevance for circular arrays of $gr$-type 
 Rydberg-atom qubits interacting through van-der-Waals type interaction outside of the Rydberg-blockade regime.

 By employing two different analog global-control-based state-preparation schemes -- respectively based on 
 NMR-type pulse sequences and shaped, time-dependent control pulses -- and making use of the dihedral symmetry of the envisioned system, we demonstrated that the two relevant classes of entangled multiqubit states
 can be engineered in the system under consideration in a time-efficient and robust fashion. More precisely, we showed that $W$ states can be engineered on time scales that are linear in the number of qubits. At 
 the same time, we found that 
 the preparation of Dicke states is feasible on time scales that scale superlinearly with the number
 of qubits; this superlinear scaling is in keeping 
 with the scalings found for Dicke state-preparation 
 schemes in other classes of physical systems. 

 Using typical values of parameters characterizing 
 realistic neutral-atom systems in optical tweezers, 
 we demonstrated that our two control schemes make it
 possible to realize $W$ and Dicke states in those systems in times much shorter than their relevant coherence times. Experimental realization of our proposed state-preparation schemes is keenly anticipated.

\appendix
\section{Further symmetries of the system}\label{app:more-symmetries}
We have introduced in Sec.~\ref{SymmSystem} the dihedral symmetry, i.e. the symmetry of the polygon with $N$ sides, as the symmetry of the system that we are concerned with in this work. However, this symmetry is, strictly
speaking, not the only symmetry of the system at hand. Let us consider the following $x$- and $y$-parity operators
\begin{align}
    X_P&=\bigotimes_{n=1}^N\sigma_n^x=\sigma_1^x\dots\sigma_N^x\\
    Y_P&=\bigotimes_{n=1}^N\sigma_n^y=\sigma_1^y\dots\sigma_N^y\ .
\end{align}
It is straightforward to check that both of these operators commute with $H_{ZZ}$, and that $H_x\:(H_y)$ commutes with $X_P 
\:(Y_P)$ respectively.
As a consequence, the two eigensubspaces $X_P^\pm$ of $X_P$, corresponding to its eigenvalues $\pm 1$, are invariant under the action of both $H_{ZZ}$ and $H_x$.
This means that, in case of dynamics driven by the only effect of $H_x$ and $H_{ZZ}$, we can further restrict the basis to the one of the current eigensubspace. The new basis is given by symmetrical (for eigenvalue $+1$) and antisymmetrical (for eigenvalue $-1$) combinations -- with respect to $X_P$ -- of the initial basis vectors. This is mathematically expressed as 
\begin{equation}
    {\cal H}_{\cal B}^{X_p^\pm}={\rm span}\left\{\frac1{\sqrt2}(\mathbbm1\pm X_p)\ket\varphi, \ket\varphi\in\mathcal B\right\}
\end{equation}
where $\mathbbm 1$ denotes the identity matrix over $N$ qubits, and the procedure is irrelevant of the initial basis $\mathcal B$, which in our case is $\mathcal B = {\rm CB}_{D_N}$.

By analogy, the same reasoning holds for $H_y,H_{ZZ}$ with respect to $Y_P$, and the restricted basis in this latter case 
is given by
\begin{equation}
\begin{split}
    {\cal H}_{\cal B}^{Y_p^\pm}&={\rm span}\left\{\frac1{\sqrt2}(\mathbbm1\pm Y_p)\ket\varphi, \ket\varphi\in\mathcal B\right\}\\
    &={\rm span}\left\{\frac1{\sqrt2}(\mathbbm1\pm (-1)^{{\rm Ham}(\ket\varphi)} i^N X_p)\ket\varphi, \ket\varphi\in\mathcal B\right\}\\
\end{split}
\end{equation}
where ${\rm Ham}(\ket\varphi)$ denotes the Hamming weight of the state $\ket\varphi$ (i.e. the number of ones of the state written in the computational basis).

Note that, by construction, there could be non-empty intersections between the subspaces ${\cal H}^{X_p^\pm}_{\cal B}$ and ${\cal H}^{Y_p^\pm}_{\cal B}$, being irrelevant the presence of any further symmetry (e.g. due to $N$ being even/odd).

\section{Numerical implementation of 
the NMR-like control scheme} \label{app:CodeTechn}
Here we provide some additional technical details about the numerical implementation
of the NMR-like control scheme presented in Sec.~\ref{sec:nmr}.

We recall first that the time-evolution
operator is parametrized in the form of Eq.~\eqref{defTotalU}
\begin{equation}
U=U_{xy}(\alpha_{M+1},\phi_{M+1}) 
\prod_{j=M}^1 U_{ZZ}(\xi_j)\:
U_{xy} (\alpha_j,\phi_j)
\end{equation}
with optimization parameters $\{\alpha_j,\phi_j,\xi_j|j=1,\dots,M\}\cup\{\alpha_{M+1},\phi_{M+1}\}$, and that we are aiming at maximizing the target-state fidelity, defined in Eq.~\eqref{eq:fidelity}.

Due to the periodicity of the operators $U_{xy}(\alpha_j,\phi_j)$ and $U_{ZZ}(\xi_j)$, we imposed appropriate box-constraints on the optimization parameters. In particular, it is easy to check that (up to global phases) the following relations hold:
\begin{align}
    U_{xy}(\alpha_j,\phi_j)&=U_{xy}(\alpha_j+\pi,\phi_i)\\
    U_{xy}(\alpha_j,\phi_j)&=U_{xy}(\alpha_j,\phi_i+2\pi)\\
    U_{ZZ}(\xi_j)&=\begin{cases}
        U_{ZZ}(\xi_j+\pi)&\text{ if $N<3$}\\
        U_{ZZ}(\xi_j+\pi/2)&\text{ if $N\geq3$}\\
    \end{cases}
\end{align}
hence the parameters have been constrained to belong to their respective intervals $\alpha_j\in[0,\pi)$, $\phi_j\in[0,2\pi)$, $\xi_j\in[0,\pi)$ if $N<3$ or $\xi_j\in[0,\pi/2)$ if $N\geq3$.
Random samples for initial guesses have been drawn from uniform distribution on the same region.

In order to maximize the efficiency, the propagator has been applied iteratively [i.e. one step $U_l$ ($l=xy,ZZ$) at a time] to the state vector of the system using the Krylov method~\cite{saad_analysis_1992};
this method allows to compute the matrix exponential applied to a vector $e^A v$ without building the whole matrix $e^A$.

The optimization is then performed through the L-BFGS minimization algorithm~\cite{nocedal_updating_1980} using the infidelity $1-{\cal F}_{t=t_f}$ as the cost function.

The code has been written in \textsc{Julia}.

\section{Numerical implementation of 
the time-dependent control scheme}
\label{app:CodeTimeDep}
Here we provide some additional technical details about the numerical implementation of the time-dependent control scheme presented in Sec.~\ref{sec:tdep}.

We recall that the time-evolution operator in this case is given by the time-ordered exponential [cf. Eq.~\eqref{eq:tdep-U}]
\begin{equation}
U(t)={\cal T}\exp\left\{-i\int_{t_i}^t H(t') {\rm d}t'\right\} \:,
\end{equation}
where $H(t)$ is the time-dependent Hamiltonian introduced in Eqs.~(\ref{eq:ham},\ref{DriftHam},\ref{ControlHam})
\begin{equation}
H(t)=H_{ZZ}+B_x(t)H_x+B_y(t)H_y
\end{equation}
and the fields are parametrized as in Eq.~\eqref{eq:sine-pulse}
\begin{equation}
B_{x,y}(t)=b^{(x,y)}_0+
\sum_{m=1}^Mb^{(x,y)}_m\sin(m\pi t/T)\:,
\end{equation}
hence the optimization parameters are $\{b_m^{(x)},b_m^{(y)} | m=0,\dots,M\}$.

In this case, given that our time-evolution operator is not discretized 
in a finite number of constant pieces [cf. Appendix~\ref{app:CodeTechn}], we cannot just iteratively apply many partial evolutions to the state vector up to the final state.
Even if there are many techniques for approximating a continuous Hamiltonian with piece-wise constant steps, e.g. 
Suzuki-Trotter decomposition~\cite{suzuki_generalized_1976}, here we have directly addressed 
the time-dependent Schr\"odinger equation
\begin{equation}
\frac{\rm d}{{\rm d}t}\ket{\psi(t)}=-iH(t)\ket{\psi(t)}
\end{equation}
and integrated it with
an adaptive explicit 5\textsuperscript{th}-order Runge–Kutta method (embedded 5(4) pair)~\cite{tsitouras_runge_2011} with $\ket{\psi(t_i=0)}=\ket{00\dots0}$ 
being the initial condition.
Given that this integration method is not designed specifically for quantum systems -- hence it is not numerically guaranteed that the norm of the final vector is preserved as in some other methods (like, for instance, the ones using matrix exponentials) -- we also applied a normalization step to the final state in order to have a well-defined state vector for computing the overlap with the target state [required for evaluating the target-state fidelity based on Eq.~\eqref{eq:fidelity}], namely $\ket{\psi(t=t_f)}=\psi_{\rm RK}(t_f)/\lVert{\psi_{\rm RK}(t_f)}\rVert$, ${\psi_{\rm RK}(t_f)}$ denoting the output vector given by the integrator at final time $t=t_f$ and $\lVert\cdot\rVert$ the vector $L_2$-norm.

We have, moreover, enforced the constraint that the control fields remain within the interval $[0, 2\pi)$: since this cannot be straightforwardly imposed as box constraints on the optimization parameters due to their non-local effect, we include it through a penalty term in the loss function, proportional to the maximum/minimum value of the fields exceeding the allowed range.

The optimization has been performed with the L-BFGS~\cite{nocedal_updating_1980} minimization algorithm, using the infidelity $1-{\cal F}_{t=t_f}$ as the cost function, together with the penalty terms. The whole loss function then reads
\begin{equation}
\begin{split}
        1&-{\cal F}+\\
         &+\Theta(B_x-2\pi)\max B_x+\Theta(B_y-2\pi)\max B_y+\\
         &-\Theta(-B_x)\min B_x-\Theta(-B_y)\min B_y
\end{split}
\end{equation}
where $\Theta$ denotes the Heaviside function ensuring that the penalty terms are effective only outside of the desired region.

All numerical routines have been implemented with \textsc{Julia} programming language.

\end{document}